\documentclass[10pt]{article} 
\usepackage[affil-it]{authblk}
\usepackage[]{graphicx}
\usepackage{amsfonts}
\usepackage{amsmath}
\usepackage{natbib}

\DeclareMathOperator*{\argmin}{arg\,min}
\newcommand{\diag}{\mathop{\mathrm{diag}}}

\title{Model-based iterative reconstruction for \\spectral-domain optical coherence tomography} 
\date{January 23 2018}
\author{Jonathan H. Mason\textsuperscript{1}}\author{Yvonne Reinwald\textsuperscript{2}}\author{Ying Yang\textsuperscript{3}}\author{Sarah Waters\textsuperscript{4}}\author{Alicia El Haj\textsuperscript{5}}\author{Pierre O. Bagnaninchi\textsuperscript{1}}
\affil{\textsuperscript{1}MRC Centre for Regenerative Medicine, The University of Edinburgh, 5 Little France Drive, Edinburgh, United Kingdom}
\affil{\textsuperscript{2}Department of Engineering, College of Science and Technology, Nottingham Trent University, Nottingham, United Kingdom}
\affil{\textsuperscript{3}Institute for Science and Technology in Medicine, Keele University, Keele, United Kingdom}
\affil{\textsuperscript{4}Mathematical Institute, University of Oxford, Oxford, United Kingdom}
\affil{\textsuperscript{5}Healthcare Technologies Institute, University of Birmingham, Birmingham, United Kingdom}

\begin{document} 
\maketitle 

\begin{abstract}
	Spectral domain optical coherence tomography (OCT) offers high resolution multidimensional imaging, but generally suffers from defocussing, intensity falloff and shot noise, causing artifacts and image degradation along the imaging depth. In this work, we develop an iterative statistical reconstruction technique, based upon the interferometric synthetic aperture microscopy (ISAM) model with additive noise, to actively compensate for these effects. For the ISAM re-sampling, we use a non uniform FFT with Kaiser-Bessel interpolation, offering efficiency and high accuracy. We then employ an accelerated gradient descent based algorithm, to minimize the negative log-likelihood of the model, and include spatial or wavelet sparsity based penalty functions, to provide appropriate regularization for given image structures. We evaluate our approach with titanium oxide micro-bead and cucumber samples with a commercial spectral domain OCT system, under various subsampling regimes, and demonstrate superior image quality over traditional reconstruction and ISAM methods.
\end{abstract}

\section{INTRODUCTION}
Optical coherence tomography offers rapid non-invasive imaging at a high resolution, and has many applications in medicine and materials science. Spectral domain OCT (SD-OCT) is a popular modality, whose spatial resolution and depth are limited by its photo-detector array and measurement noise, especially when direct reconstruction is used. With model-based iterative reconstruction (MBIR), which is actively researched in medical imaging, one instead takes an optimization approach to forming an image. One searches for the statistically most likely specimen structure, given a set of noisy or incomplete measurements, according to a defined system model and exploiting a priori information.

A similar instance of MBIR is compressive SD-OCT proposed by Liu and Kang\cite{Liu2010}. This minimized an objective function combining a least squares penalty term with $\l_1$ regularization to randomly sub-sampled measurements from a common path set-up. This was then extended by Xu et al.\cite{Xu2013} to include dispersion correction. Additionally, their model implicitly assumed that A-scans are independent, which will not account for out of focus blurring by the optics.

It was discovered by Ralston et al.\cite{Ralston2007} that the spectral domain representations of an object and its OCT measurements are connected by means of a re-sampling and filtering. With this, they proposed synthetic aperture interferometric microscopy (ISAM) as a direct regularized inversion of this model, which achieves a refocusing of each layer of the specimen.

In this article, we propose a MBIR approach for SD-OCT, which combines the ISAM model with dispersion correction, and can incorporate appropriate regularization based upon the spatial structure of the specimen. To this end, we propose an efficient algorithm, permitting many regularization functions.

\section{METHODOLOGY}
In this section, we will detail the model and optimization algorithm used for OCT reconstruction. 
\subsection{Model Definition}
A general form of MBIR is solution of the following objective function
\begin{equation} \label{equ:obj}
\boldsymbol{x} = \argmin_{\boldsymbol{x}} \underbrace{\mathcal{D}(\boldsymbol{x};\boldsymbol{y})}_\mathrm{data\,fidelity}+\underbrace{\lambda\mathcal{R}(\boldsymbol{x})}_\mathrm{regularization},
\end{equation}
where $\boldsymbol{x}\in\mathbb{C}^N$ is the complex valued image we wish to estimate and $\boldsymbol{y}\in\mathbb{R}^M$ are the real valued measurements from the spectrometer. The function $\mathcal{R}$ is to promote desired structure within the image according to a priori information, and the scalar $\lambda$ weights its strength. Choices for $\mathcal{R}$ are explored in Section~\ref{sec:reg}.

The first term in ($\ref{equ:obj}$), $\mathcal{D}(\cdot;\cdot)$, is a `data fidelity' term. It measures the mismatch between $\boldsymbol{x}$ and measurements $\boldsymbol{y}$, according to the system model and the statistics of the noise. In this article, we opt for the squared residual norm defined as
\begin{equation} \label{equ:ls_obj}
\mathcal{D}(\boldsymbol{x};\boldsymbol{y}) = \frac{1}{2}\|\boldsymbol{\Phi}(\boldsymbol{x})-\boldsymbol{y}\|_2^2,
\end{equation}
where $\boldsymbol{\Phi}$ is our system model operator. This choice implies additive zero-mean Gaussian noise on our measurements, which we note is an approximation to the shot noise on the detector. Advantages of using (\ref{equ:ls_obj}) are its convexity and the simplicity with which it can be minimized. This can be addressed through gradient descent, for which the derivative of (\ref{equ:ls_obj}) can be calculated as
\begin{equation} \label{equ:ls_grad}
\boldsymbol{\nabla}\mathcal{D}(\boldsymbol{x};\boldsymbol{y}) = \boldsymbol{\Phi}^*\left(\boldsymbol{\Phi}(\boldsymbol{x})-\boldsymbol{y}\right),
\end{equation}
where $\boldsymbol{\Phi}^*$ is the adjoint operator.

A key component of our system model represented by $\boldsymbol{\Phi}$ is the ISAM mapping, which is active of the spatial Fourier transform of the object denoted as
\begin{equation}
\hat{\boldsymbol{x}}(\boldsymbol{q},\boldsymbol{\beta}) \equiv \mathcal{F}(\boldsymbol{x}),
\end{equation}
where $\mathcal{F}(\cdot)$ is the Fourier transform, $\boldsymbol{q}$ are lateral frequencies, and $\boldsymbol{\beta}$ are axial frequencies in object space. The ISAM mapping is then
\begin{equation}
\hat{\boldsymbol{u}}(\boldsymbol{q},\boldsymbol{k}) = \diag\{\hat{\boldsymbol{f}}(\boldsymbol{q},\boldsymbol{k})\}\left[\hat{\boldsymbol{x}}(\boldsymbol{q},\boldsymbol{k})\right]_{\boldsymbol{k} = \frac{1}{2} \sqrt{\boldsymbol{\beta}^2+\boldsymbol{q}^2}},
\end{equation}
where $\boldsymbol{k}$ is the detected wavenumber, and  $\hat{\boldsymbol{f}}(\boldsymbol{q},\boldsymbol{k})$ is a filter\cite{Ralston2007}, which in practice has minimal effect. The key to the method is therefore a geometric resampling from $\boldsymbol{\beta}$ to $\boldsymbol{k}$.

To realise ISAM, we perform the mapping directly through the nonuniform FFT (NUFFT) by Fessler and Sutton\cite{Fessler2003a} with Kaiser--Bessel interpolation for the resampling. We represent this with the single linear operation
\begin{equation}
\boldsymbol{u} = \boldsymbol{K}\boldsymbol{x},
\end{equation}
where $\boldsymbol{u}$ is the inverse Fourier transform of $\hat{\boldsymbol{u}}$. Following the ISAM mapping, we wish to also account for the effect of dispersion and map into a real measurement, as with $\boldsymbol{y}$.

This is done by: zero-padding to the full range; taking a 1D FFT to map into k-space; applying a phase compensation for dispersion correction; and finally taking the real part. Our full model is therefore given by
\begin{equation} \label{equ:for_model}
\boldsymbol{\Phi}(\boldsymbol{x}) = \diag(\boldsymbol{s})\Re\{\diag(e^{j\boldsymbol{\Omega}})\boldsymbol{D}\boldsymbol{Q}\boldsymbol{Kx}\},
\end{equation}
where $\boldsymbol{D}\in\mathbb{R}^{2N\times 2N}$ is the 1D discrete Fourier transform (DFT) matrix, which is calculated in practice through the FFT. The vector $\boldsymbol{s}$ is a binary sub-sampling vector defining the known measurements. The padding matrix $\boldsymbol{Q}\in\mathbb{R}^{2N\times N}$ is defined as
\begin{equation}
\boldsymbol{Q} = 
\begin{bmatrix}
\boldsymbol{I} \\
\boldsymbol{0} 
\end{bmatrix},
\end{equation}
where $\boldsymbol{I}\in\mathbb{R}^{N\times N}$ is the identity matrix and $\boldsymbol{0}\in\mathbb{R}^{N\times N}$ is the all zero matrix. The dispersion correction is performed with
\begin{equation}
\boldsymbol{\Omega} = a(\boldsymbol{k}-k_0)^2+b(\boldsymbol{k}-k_0)^3,
\end{equation}
where $k_0$ is the central wavenumber of the source, and scalar parameters $a$ and $b$ are selected to correct for 2nd and 3rd order dispersion mismatch between sample and reference arms.

We calculate the adjoint operator used in (\ref{equ:ls_grad}) as
\begin{equation} \label{equ:back_model}
\boldsymbol{\Phi}^*(\boldsymbol{y}) = \boldsymbol{K}^H\boldsymbol{D}^H\boldsymbol{Q}^T\diag(e^{-j\boldsymbol{\Omega}})\mathcal{A}(\diag(\boldsymbol{s})\boldsymbol{y}),
\end{equation}
where the operator $\mathcal{A}(\cdot)$ calculates the conjugate analytic signal as
\begin{equation}
\mathcal{A}(\boldsymbol{z}) = \Re(\boldsymbol{z})-j\mathcal{H}(\boldsymbol{z}),
\end{equation}
where $\mathcal{H}(\cdot)$ is the Hilbert transform.

An interesting feature of our model definition in (\ref{equ:for_model}) is its apparent non-linearity, due to taking the real part. With this, the backward operator in (\ref{equ:back_model}) in not strictly its adjoint for a general complex argument. However, given that $\diag(e^{j\boldsymbol{\Omega}})\boldsymbol{D}\boldsymbol{Q}\boldsymbol{Kz}$ is an analytic signal, then this does hold, which is guaranteed by use of the matrix $\boldsymbol{Q}$ prior to the FFT in (\ref{equ:for_model}).

\subsection{Regularization} \label{sec:reg}
In this work, we will focus on sparsity promoting regularization functions, since these will allow us to reduce the number of sufficient measurements fo reconstruction. These take the form
\begin{equation}
\mathcal{R}(\boldsymbol{x}) = \|\boldsymbol{\Psi x}\|_1,
\end{equation}
where $\boldsymbol{\Psi}$ is a basis transform in which $\boldsymbol{x}$ is known to be sparse.

\begin{itemize}
\item If $\boldsymbol{x}$ is known to be spatially sparse then one could choose
\begin{equation}
\boldsymbol{\Phi} = \boldsymbol{I}.
\end{equation}

\item Another suitable choice may be to use a wavelet transformation for $\boldsymbol{\Psi}$. Here, we opt for the dual tree complex wavelet transform (DT-CWT)\cite{Selesnick2005}, since we found this provided good performance on our test data.

\item If $\boldsymbol{x}$ is known to be piecewise constant then one could choose the isotropic total variation (TV). In two dimensions this is given by as
\begin{equation}
\mathcal{R}(\boldsymbol{x}) = \sum_{i,j}\sqrt{|x(i+1,j)-x(i,j)|^2+|x(i,j+1)-x(i,j)|^2}.
\end{equation}
\end{itemize}

\subsection{Optimization Approach}
To minimise our objective function, which is convex but has a potentially non-smooth regularization function, there are many existing algorithms. In this work, we opt for the fast iterative shrinkage-thresholding algorithm (FISTA) \cite{Beck2009}, which is an accelerated proximal gradient descent given as
\begin{align} \label{equ:alg}
\boldsymbol{z}^{k} &:= \mathbf{prox}_{\mathcal{R},\delta}\left(\boldsymbol{x}^k-\delta\boldsymbol{\nabla}\mathcal{D}(\boldsymbol{x}^k;\boldsymbol{y})\right) \nonumber \\
t^{k+1} &:= \frac{1+\sqrt{1+4(t^k)^2}}{2} \nonumber \\
\boldsymbol{x}^{k+1} &:= \boldsymbol{z}^{k}+\frac{t^k-1}{t^{k+1}}(\boldsymbol{z}^{k}-\boldsymbol{z}^{k-1})
\end{align}

The proximal operator in (\ref{equ:alg}) is defined as
\begin{equation}
\mathbf{prox}_{\mathcal{R},\delta}(\boldsymbol{v}) = \argmin_{\boldsymbol{x}\in\mathcal{C}} \frac{1}{2\delta}\|\boldsymbol{x}-\boldsymbol{v}\|_2^2+\lambda \mathcal{R}(\boldsymbol{x}).
\end{equation}
The set $\mathcal{C}$ allows one to impose box constraints on the estimate if desired. The action of this operator is to find an $\boldsymbol{x}$ in the neighbourhood of $\boldsymbol{u}$ that has the lowest value of regularization function. The size of this search radius is also dependent on the gradient step-size $\delta$. This method can be proven to be globally convergence for our objective function in (\ref{equ:obj}), given an appropriately small $\delta$.

\section{EXPERIMENTATION}
\subsection{Data and its Processing}
The data from our testing was acquired on a commercial SD-OCT system with central wavelength of 830 nm. One sample consisted of titanium oxide micro-beads dispersed in an agarose gel with a ratio of 0.6 mg/g. Another was a section of cucumber tissue. In both cases we collected a 3D volume of $1024 \times 1024$ A-scans over a square area of 4 mm\textsuperscript{2}, and extracted raw data from the spectrometer. We manually adjusted the focal plane to be positioned in the center of each specimen. For background correction, we subtracted the smoothed average along these A-scans.

\begin{figure}[htb!]
	\begin{center}
		\begin{tabular}{c|c|c|c}
			&IFFT & dispersion corrected & ISAM\\
			\hline
			\rotatebox{90}{beaded gel}&
			\includegraphics[height=4cm]{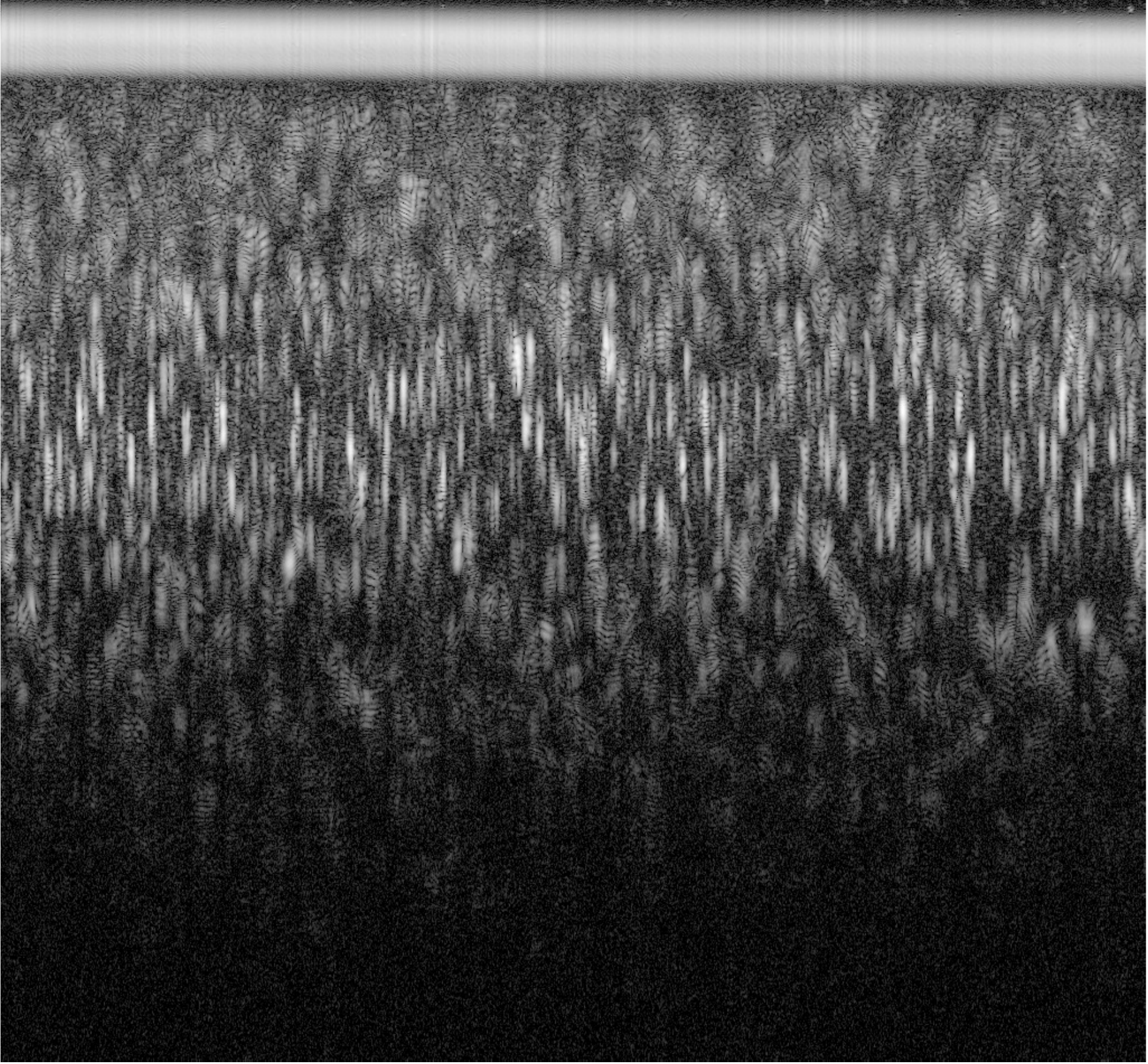}&
			\includegraphics[height=4cm]{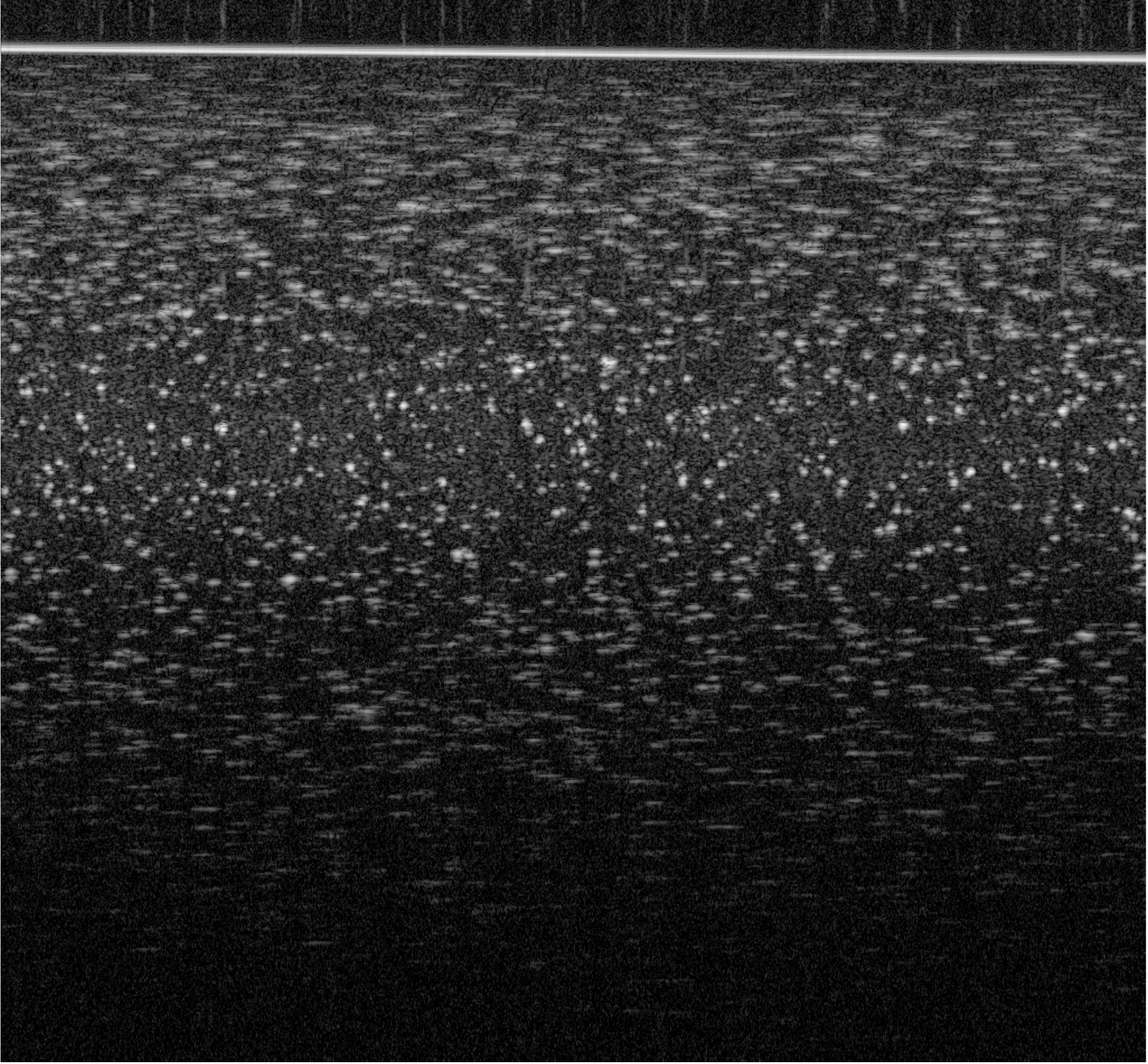}&
			\includegraphics[height=4cm]{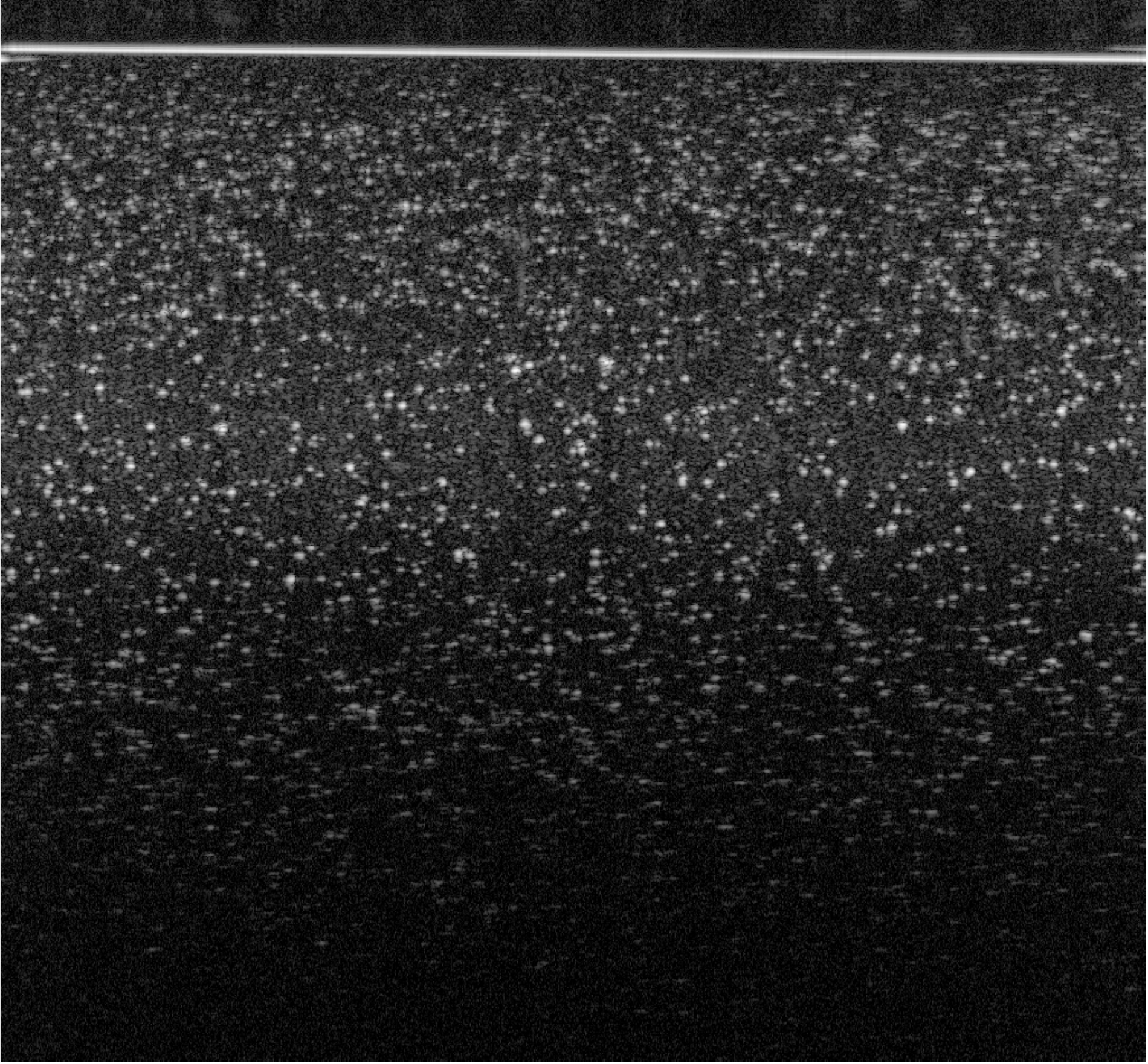}\\
			\hline
			\rotatebox{90}{cucumber}&
			\includegraphics[height=4cm]{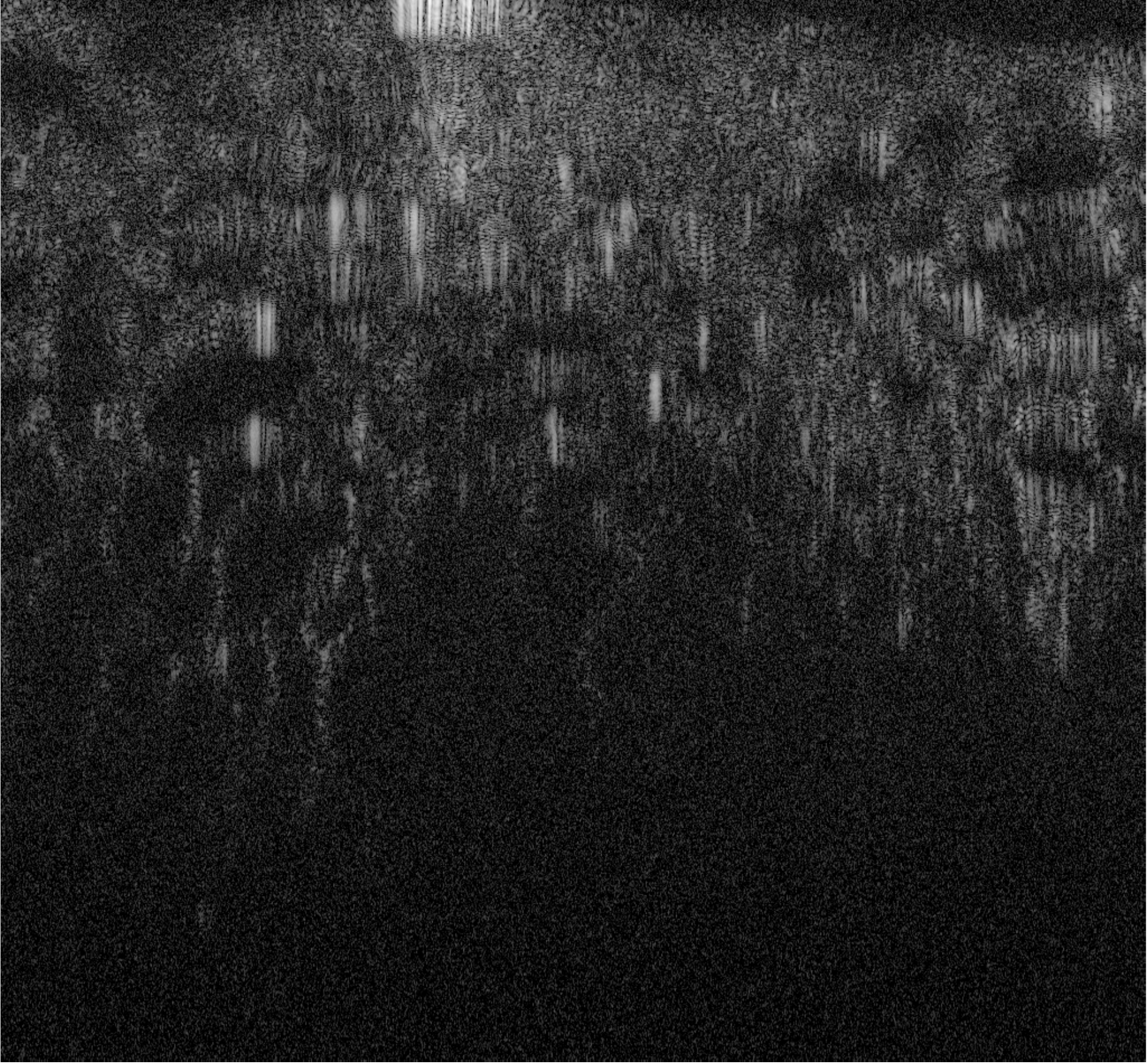}&
			\includegraphics[height=4cm]{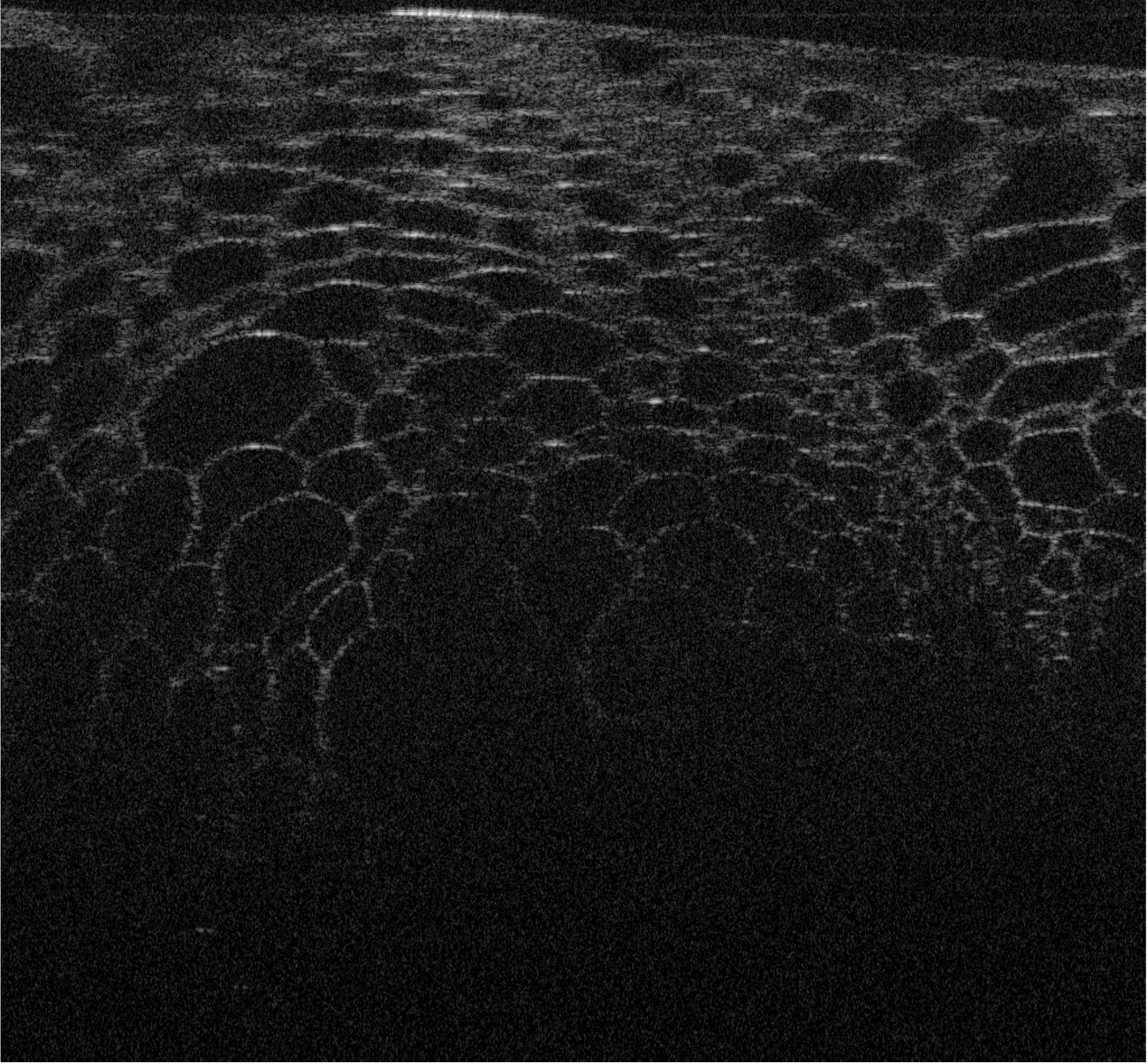}&
			\includegraphics[height=4cm]{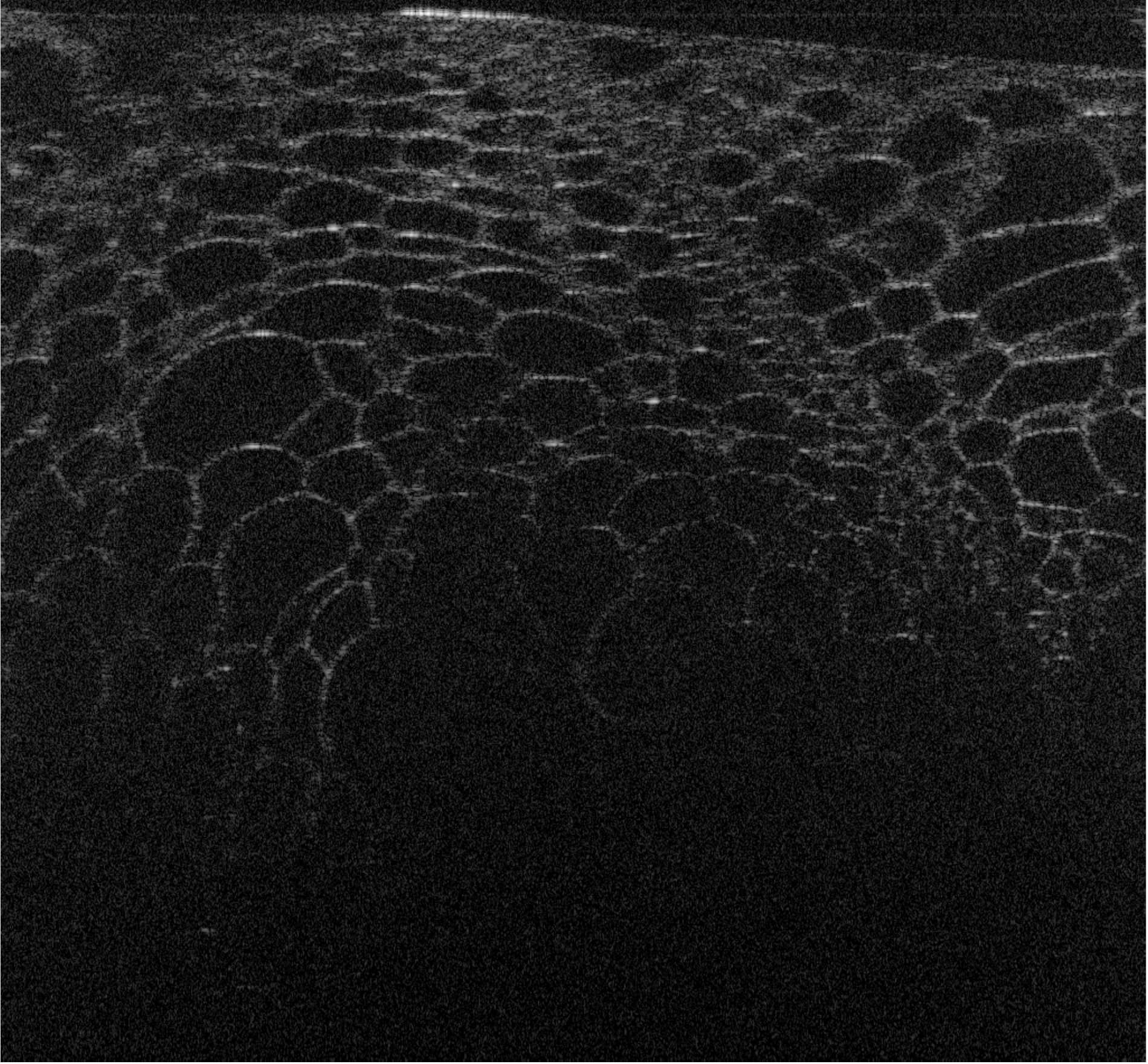}\\
			\hline
			
		\end{tabular}
	\end{center}
	\caption[example] 
	{\label{fig:samples} Samples used for testing, and various reconstructions from the full set of measurements. The left column are from IFFT with no dispersion correction; the central column with dispersion compensation and IFFT; and the right column with dispersion compensation and ISAM.}
\end{figure}

Cross sections from the samples are shown in Figure~\ref{fig:samples}. This highlights the impact of dispersion compensation, and the increase in lateral resolution away from the focal plane with the ISAM method.

We performed MBIR on a desktop PC, with an implementation of (\ref{equ:alg}) written in Matlab. We ran the following on sub-sampling patterns for $\boldsymbol{s}$:
\begin{itemize}
	\item \textit{Random}: setting the elements of $\boldsymbol{s}$ randomly with binomial distribution with expectation 0.5, halving the effective number of samples. This is the case studied in compressive SD-OCT\cite{Liu2010}.
	\item \textit{Equispaced}: the spectrometer measurements were uniformly sub-sampled in wavelength, again halving the number of samples. This would normally have the effect of reducing the imaging depth of the system.
	\item \textit{Partial}: a fully sampled sub-section of the spectrometer was selected, with half the wavelength range taken the center of the full range. This reduces the effect bandwidth of the sensor, and would normally decrease the axial resolution.
\end{itemize}
In each of the sub-sampling cases we will evaluate the performance of MBIR with TV and DT-CWT regularization functions. As a baseline, we also evaluate the result of direct ISAM on linearly interpolated measurements.
\subsection{Evaluation Method}
Each method was run for 100 iterations, which was deemed sufficient to ensure convergence in each case tested. The quality assessment metric used is the normalized cross-correlation (NCC) of magnitudes defined as
\begin{equation} \label{equ:ncc}
r(\boldsymbol{x}) = \frac{\sum_{i=1}^{N}(|x_i|-\bar{x})(|z_i|-\bar{z})}{\sqrt{\sum_{i=1}^{N}(|x_i|-\bar{x})^2}\sqrt{\sum_{i=1}^{N}(|z_i|-\bar{z})^2}},
\end{equation}
where $\boldsymbol{z}$ is a ground truth and $\bar{z}$ is its mean magnitude value. The motivation of using correlation instead of the more commonly used root--mean--squared--error (RMSE), is due to the scaling from the ISAM method.

For the regularization parameter $\lambda$, this was optimised to maximise the NCC, by means of a golden line search for each case.

\subsection{Results}
\begin{figure}[htb!]
	\begin{center}
		\begin{tabular}{c|c|c|c|}
			&interpolated ISAM & MBIR TV & MBIR DT-CWT\\
			\hline
		    \rotatebox{90}{random}&
		    \includegraphics[height=4cm]{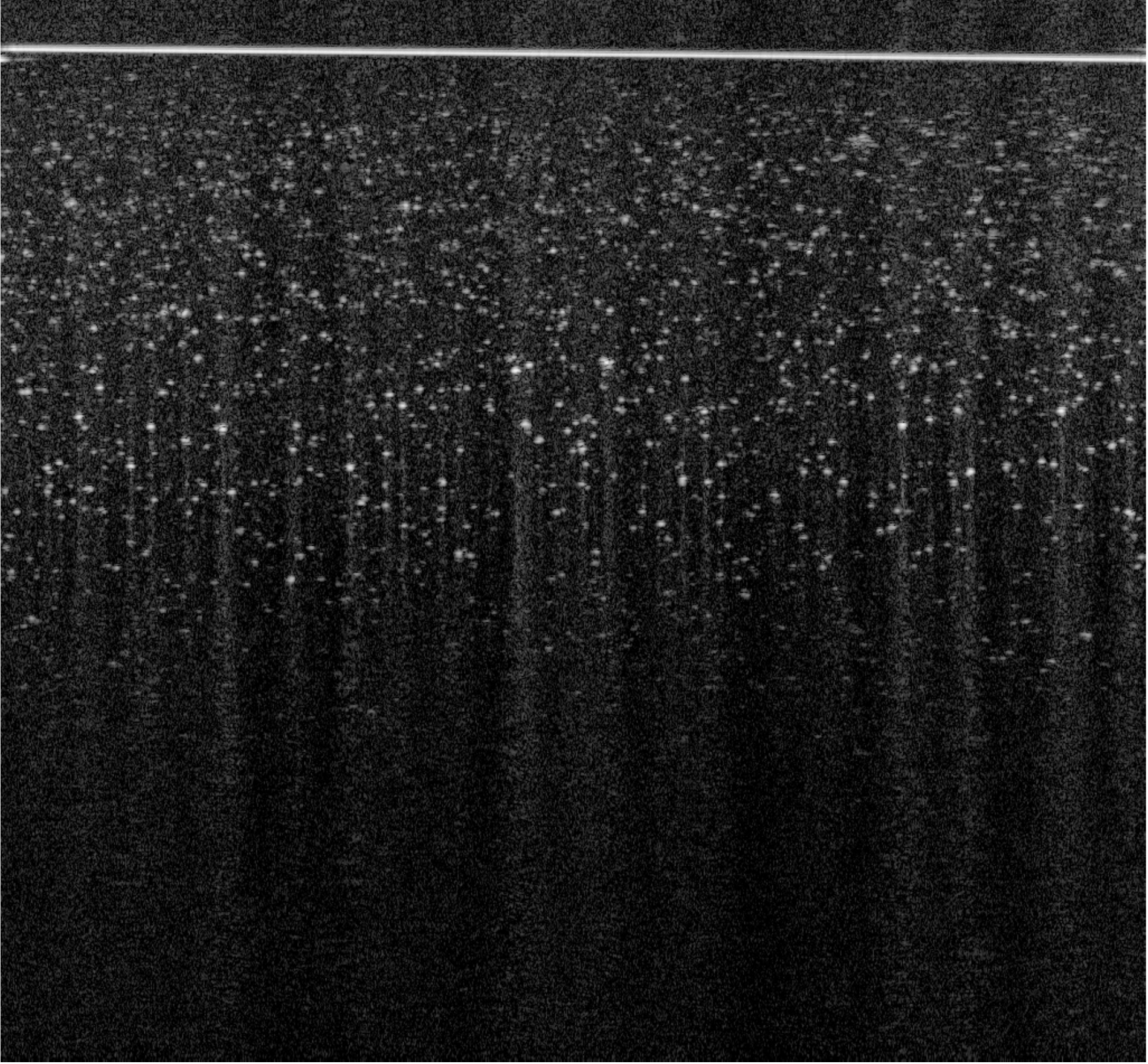}&
			\includegraphics[height=4cm]{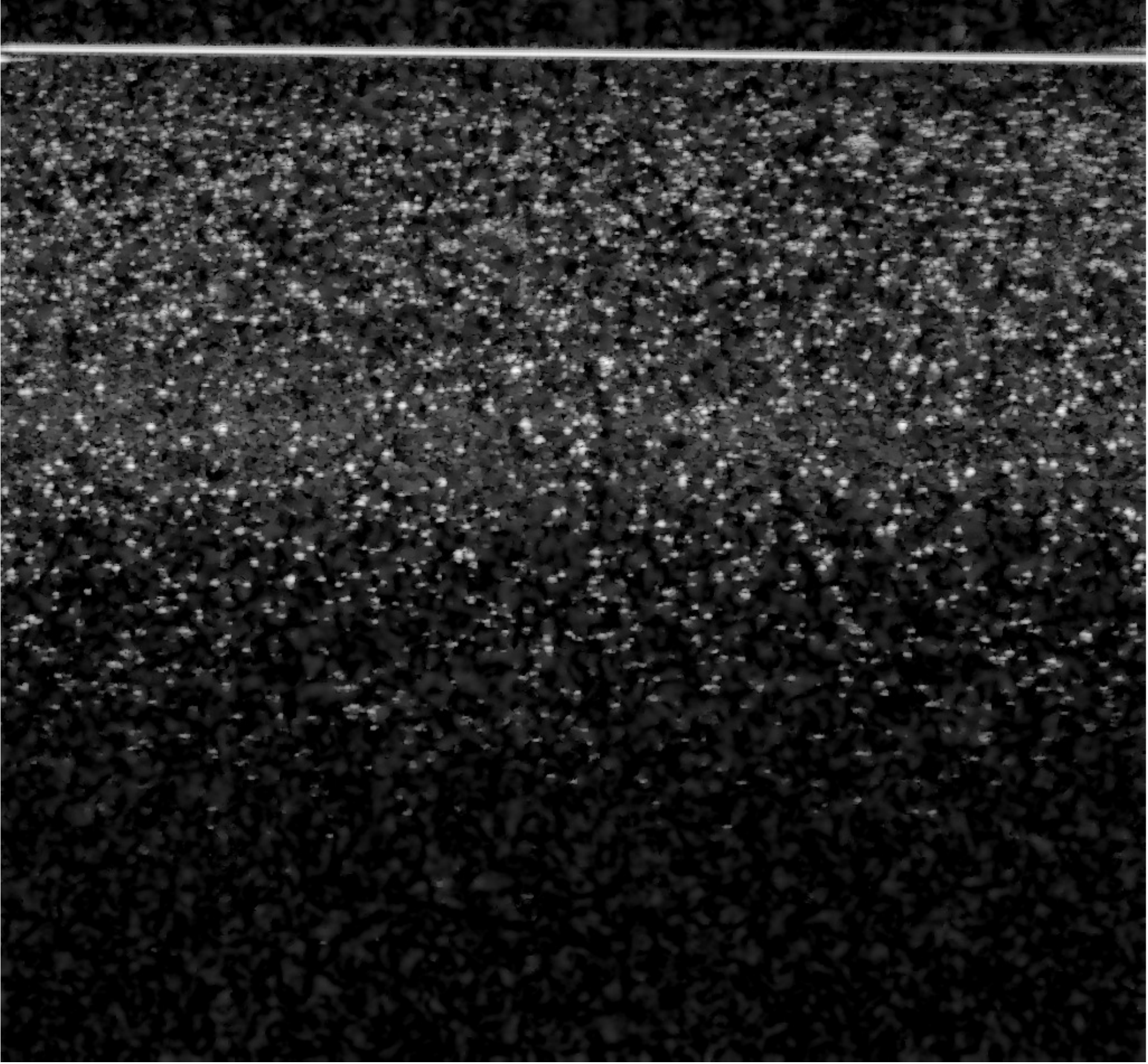}&
			\includegraphics[height=4cm]{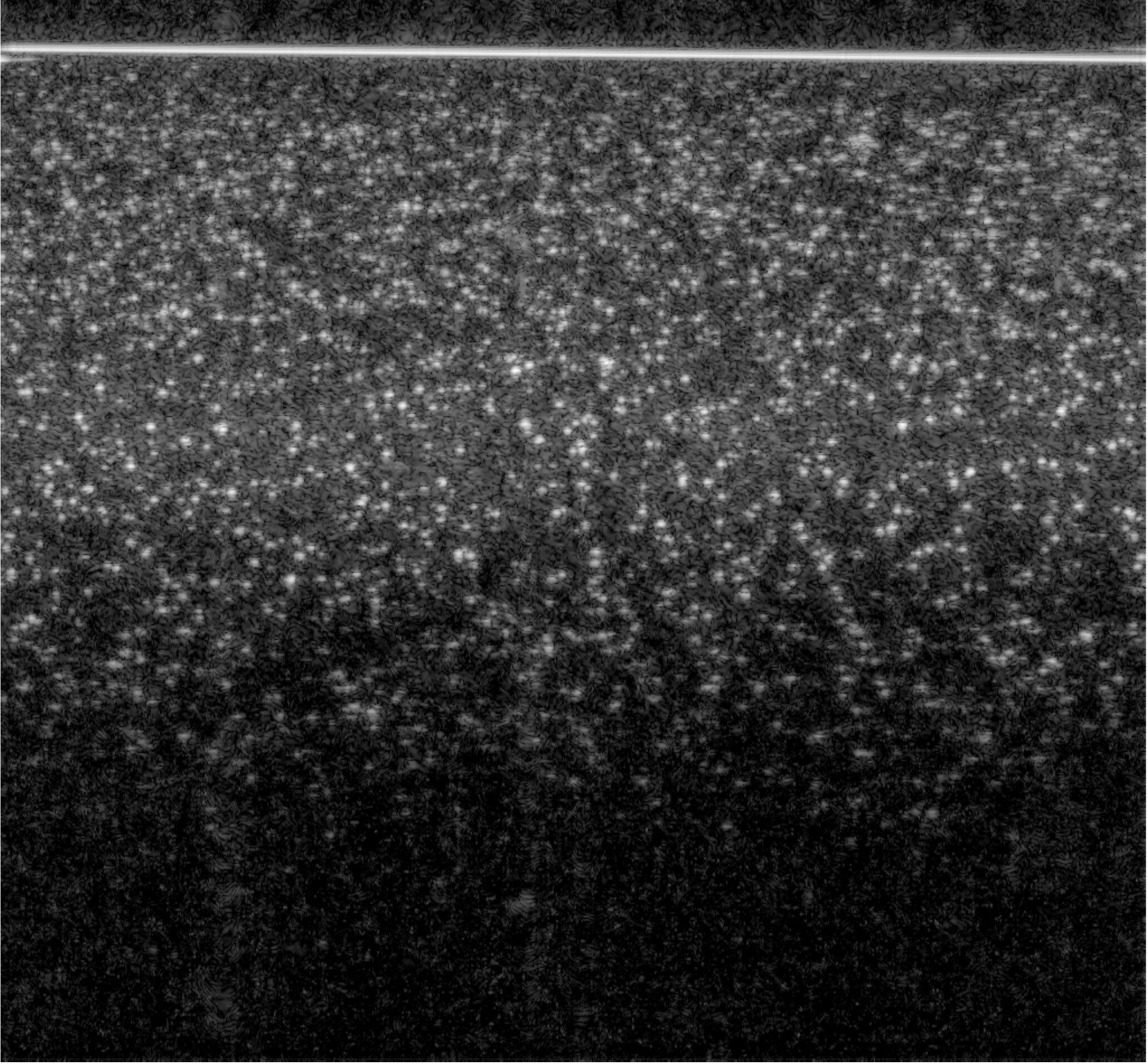}\\
			\hline
			\rotatebox{90}{equispaced}&
			\includegraphics[height=4cm]{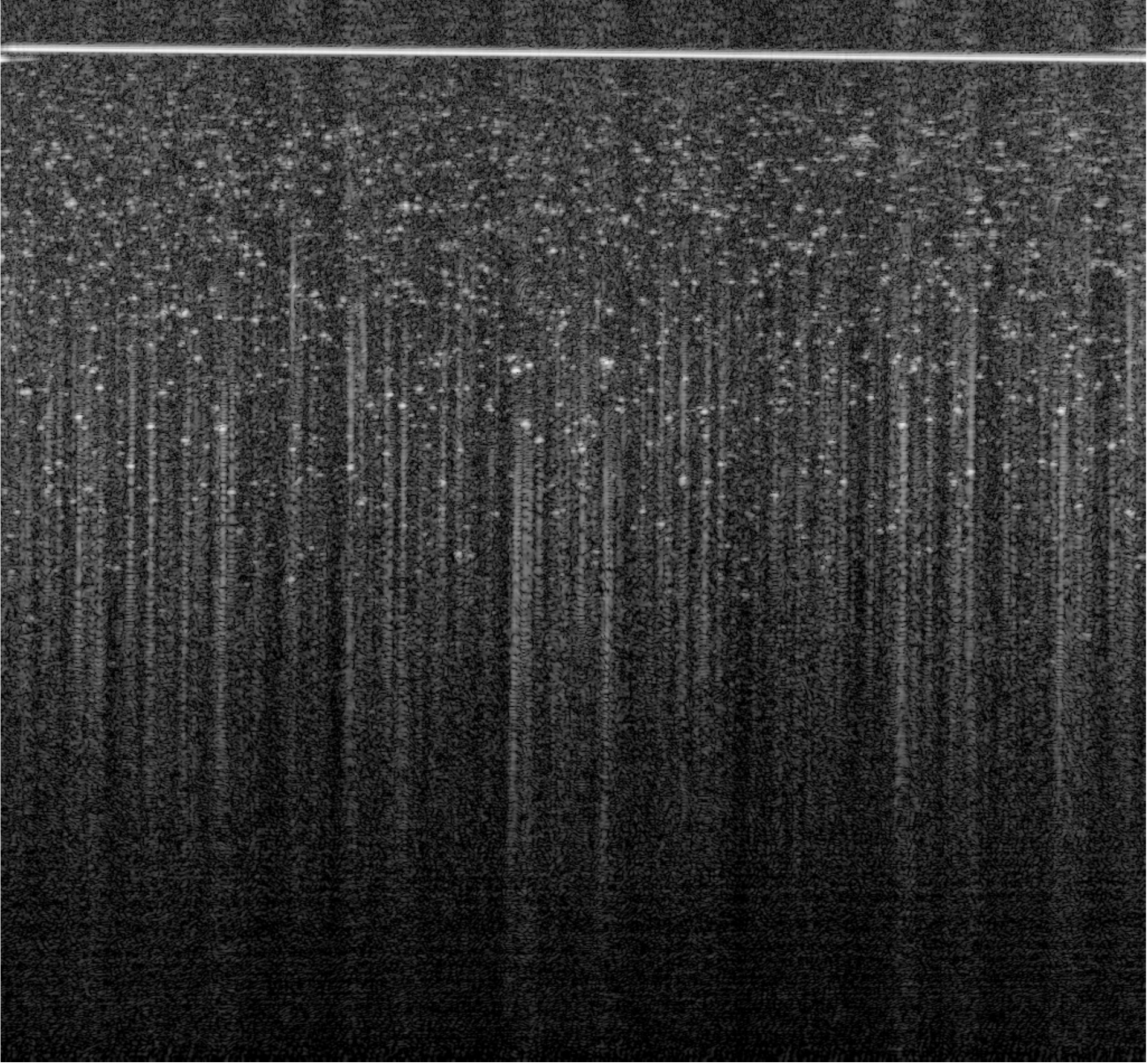}&
			\includegraphics[height=4cm]{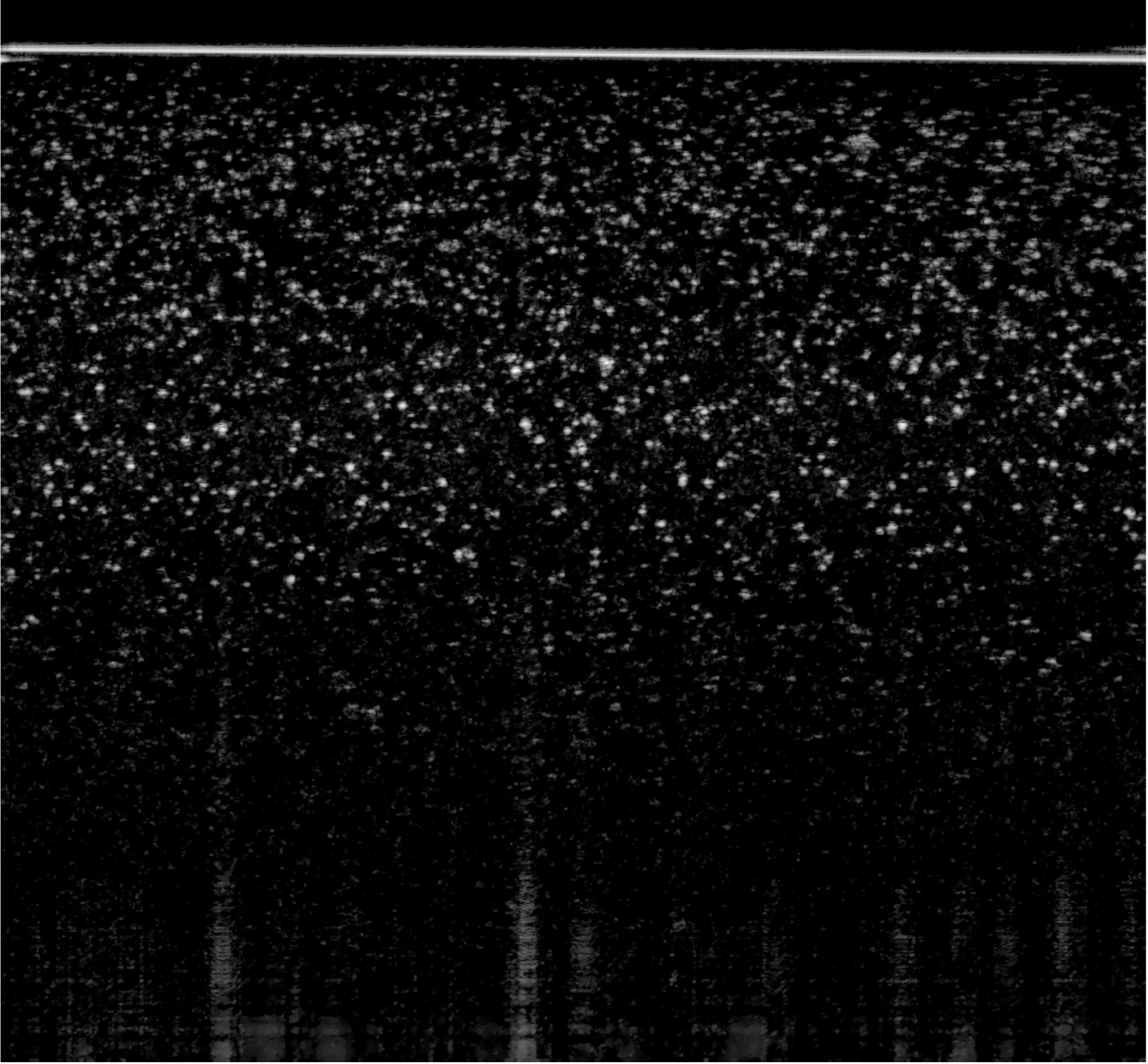}&
			\includegraphics[height=4cm]{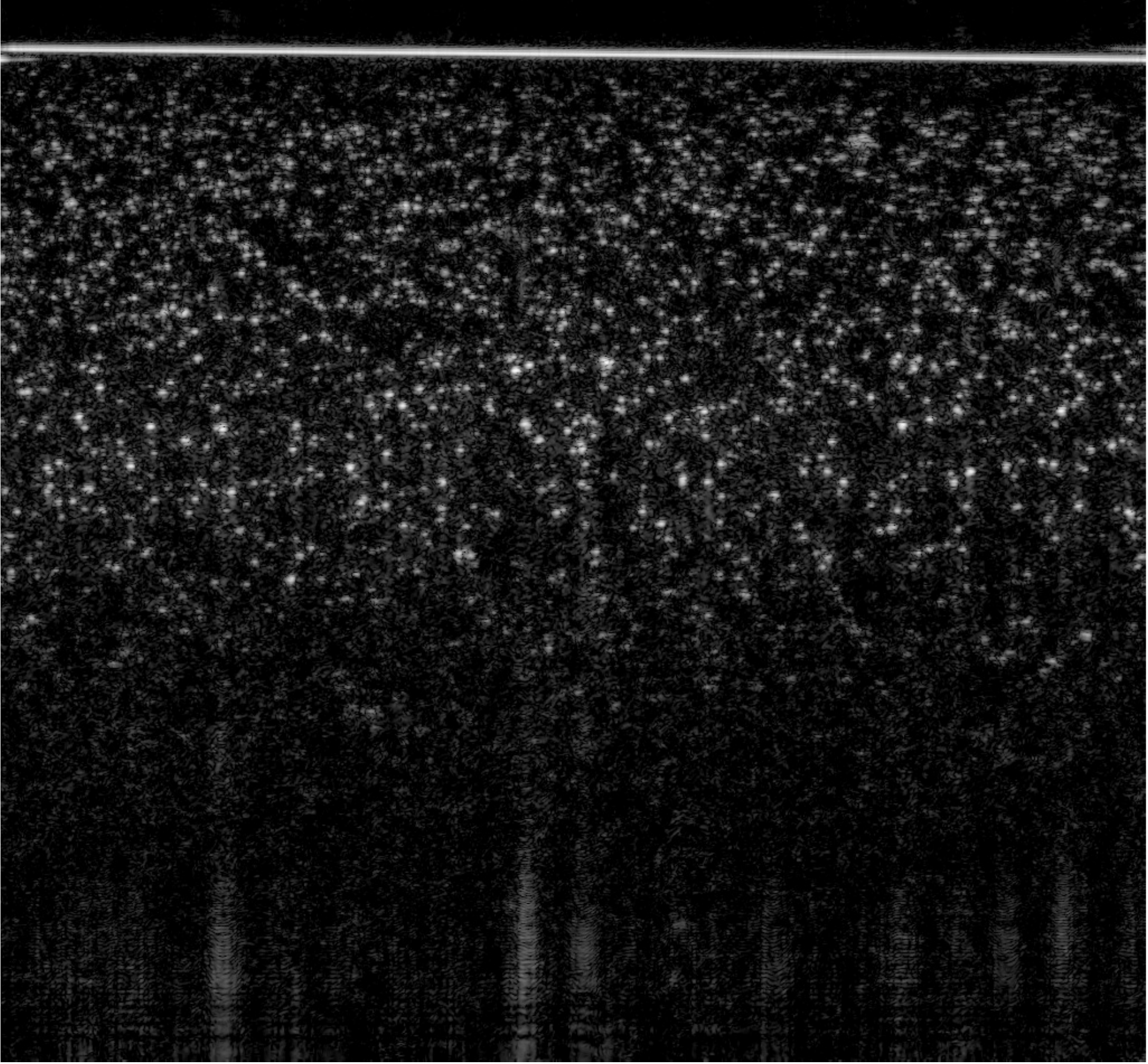}\\
			\hline
			\rotatebox{90}{partial}&
			\includegraphics[height=4cm]{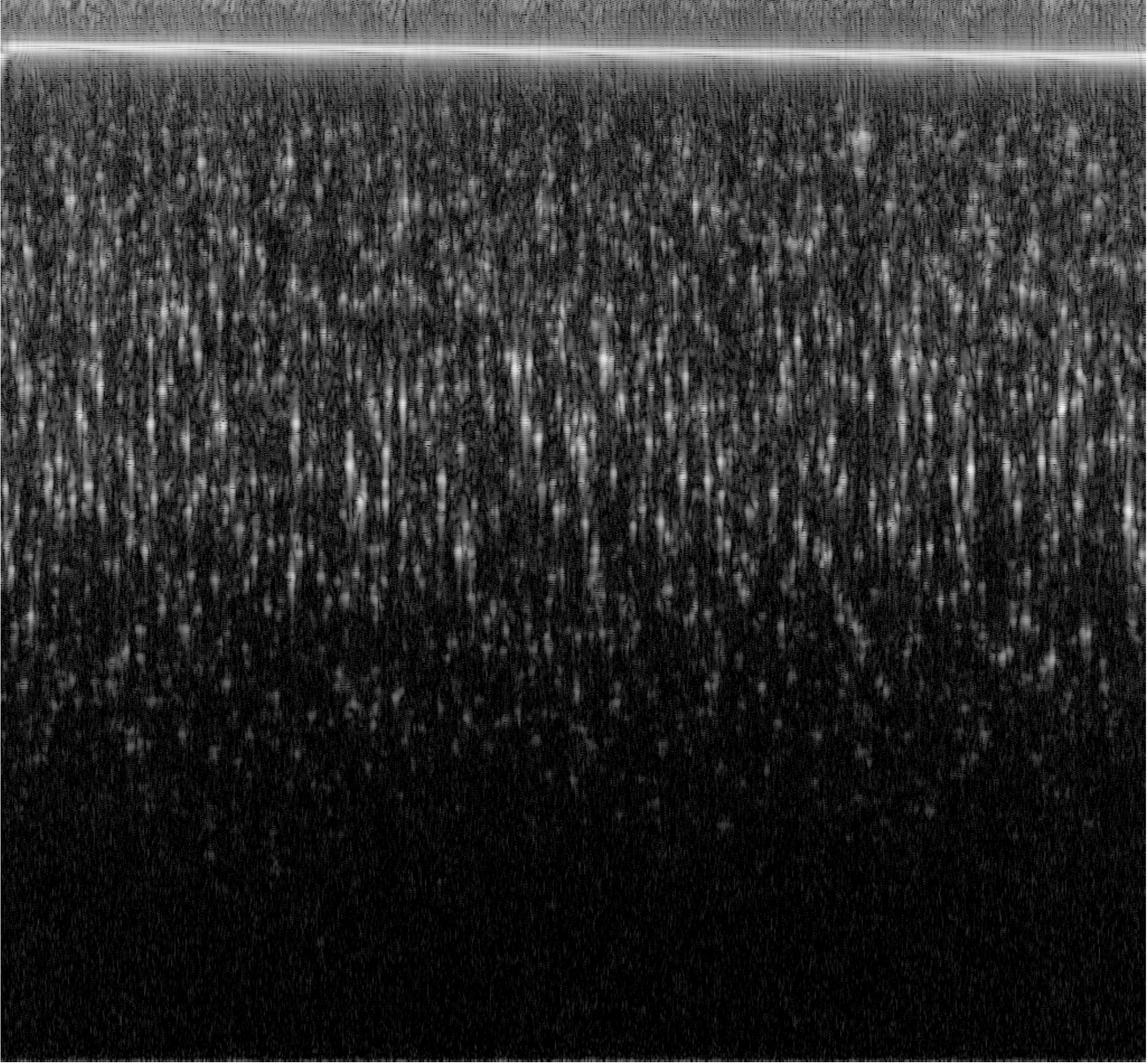}&
			\includegraphics[height=4cm]{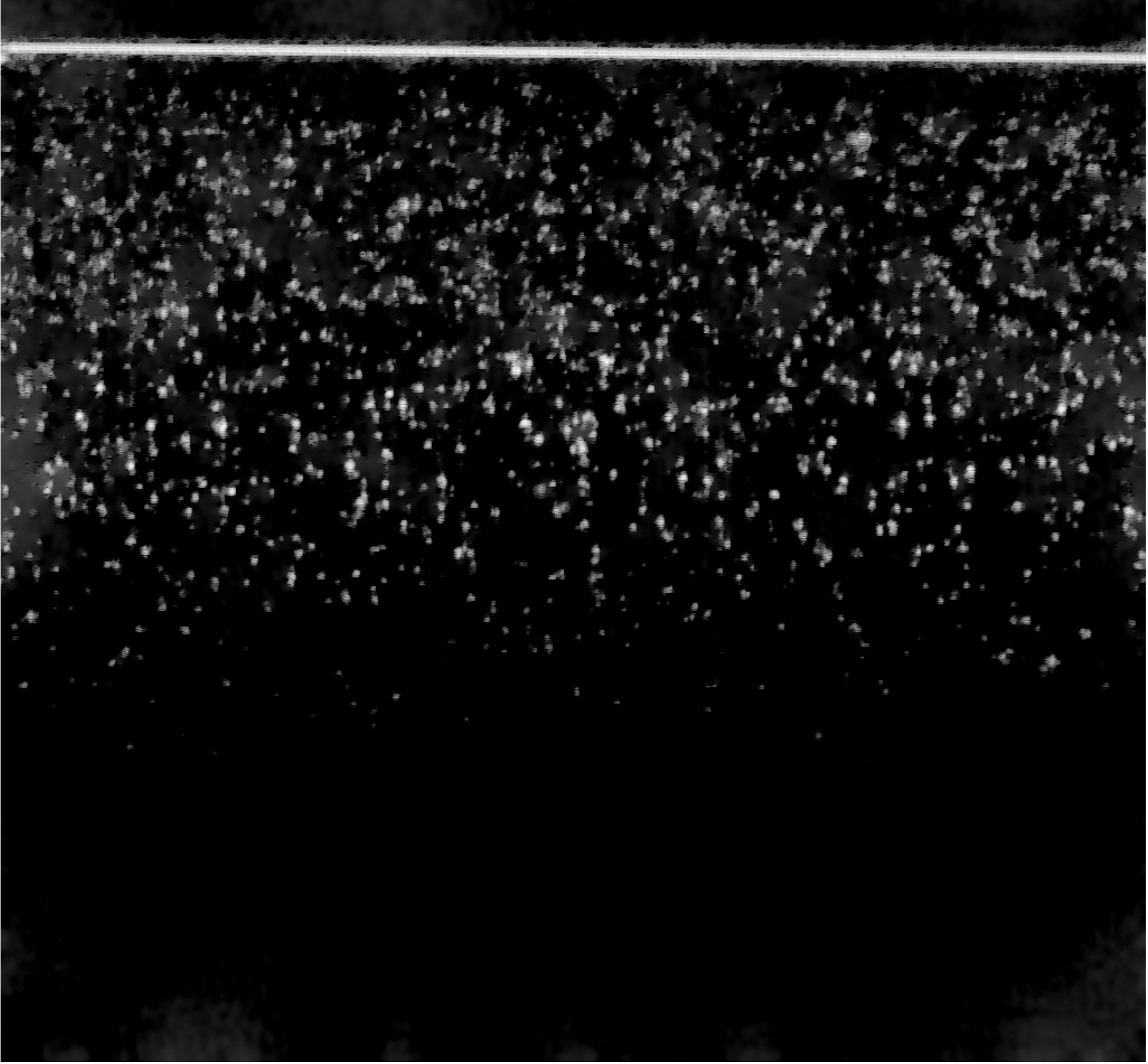}&
			\includegraphics[height=4cm]{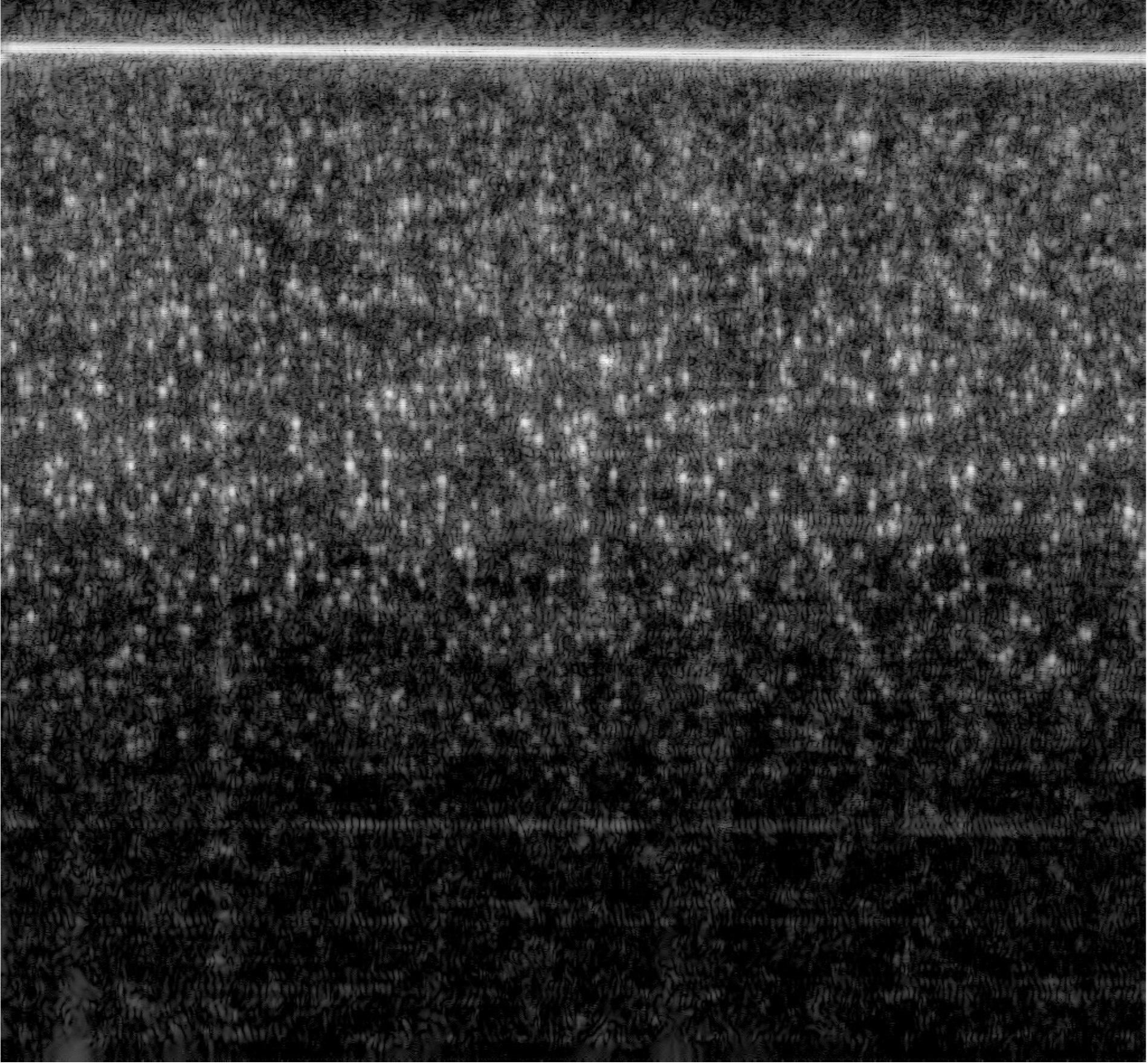}\\
			\hline
		\end{tabular}
	\end{center}
	\caption[example] 
	{\label{fig:beads} Visual results of beaded gel phantom reconstructions from various sub-sampling schemes.}
\end{figure}

\begin{figure}[htb!]
	\begin{center}
		\begin{tabular}{c|c|c|c|}
			&interpolated ISAM & MBIR TV & MBIR DT-CWT\\
			\hline
			\rotatebox{90}{random}&
			\includegraphics[height=4cm]{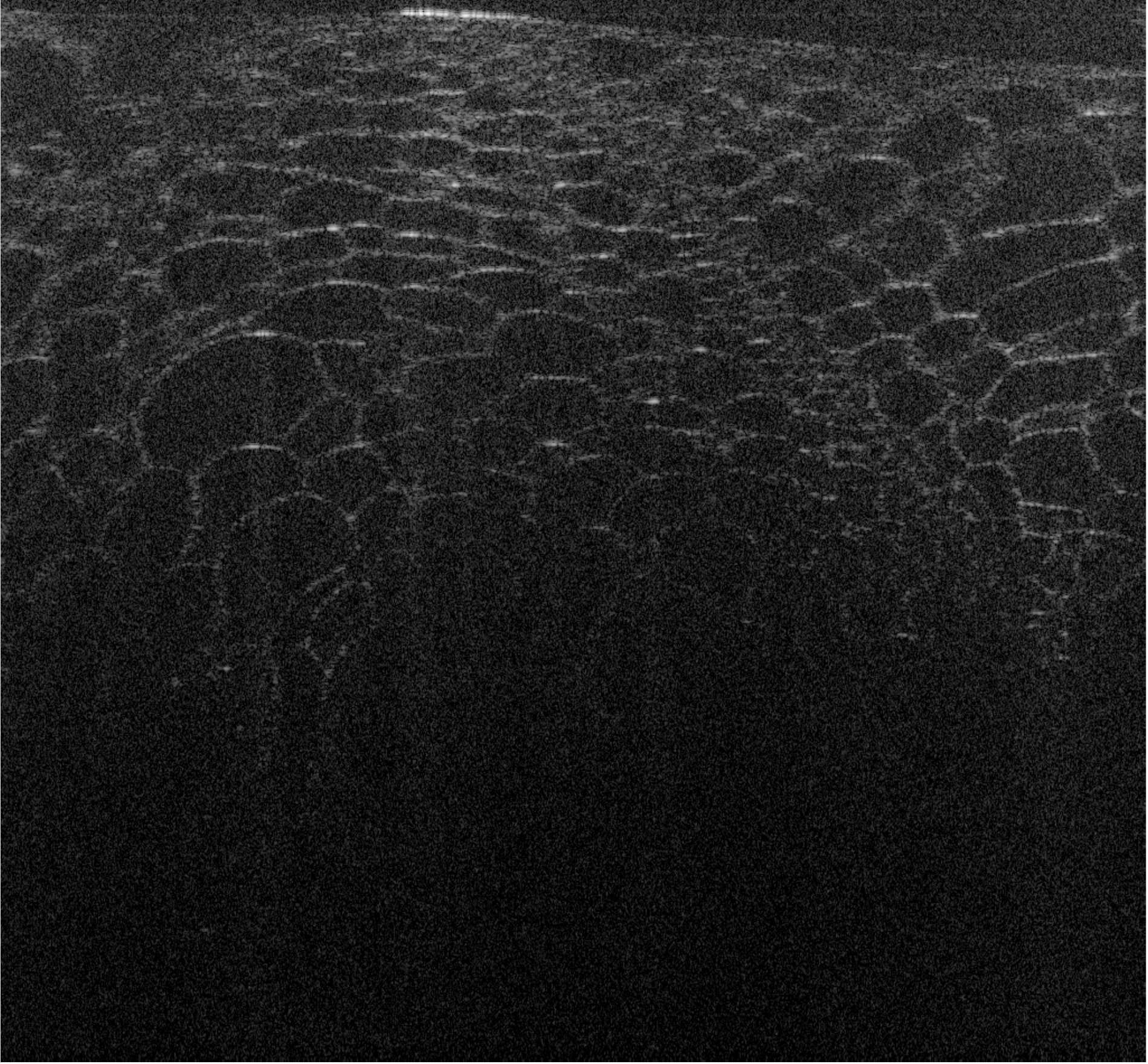}&
			\includegraphics[height=4cm]{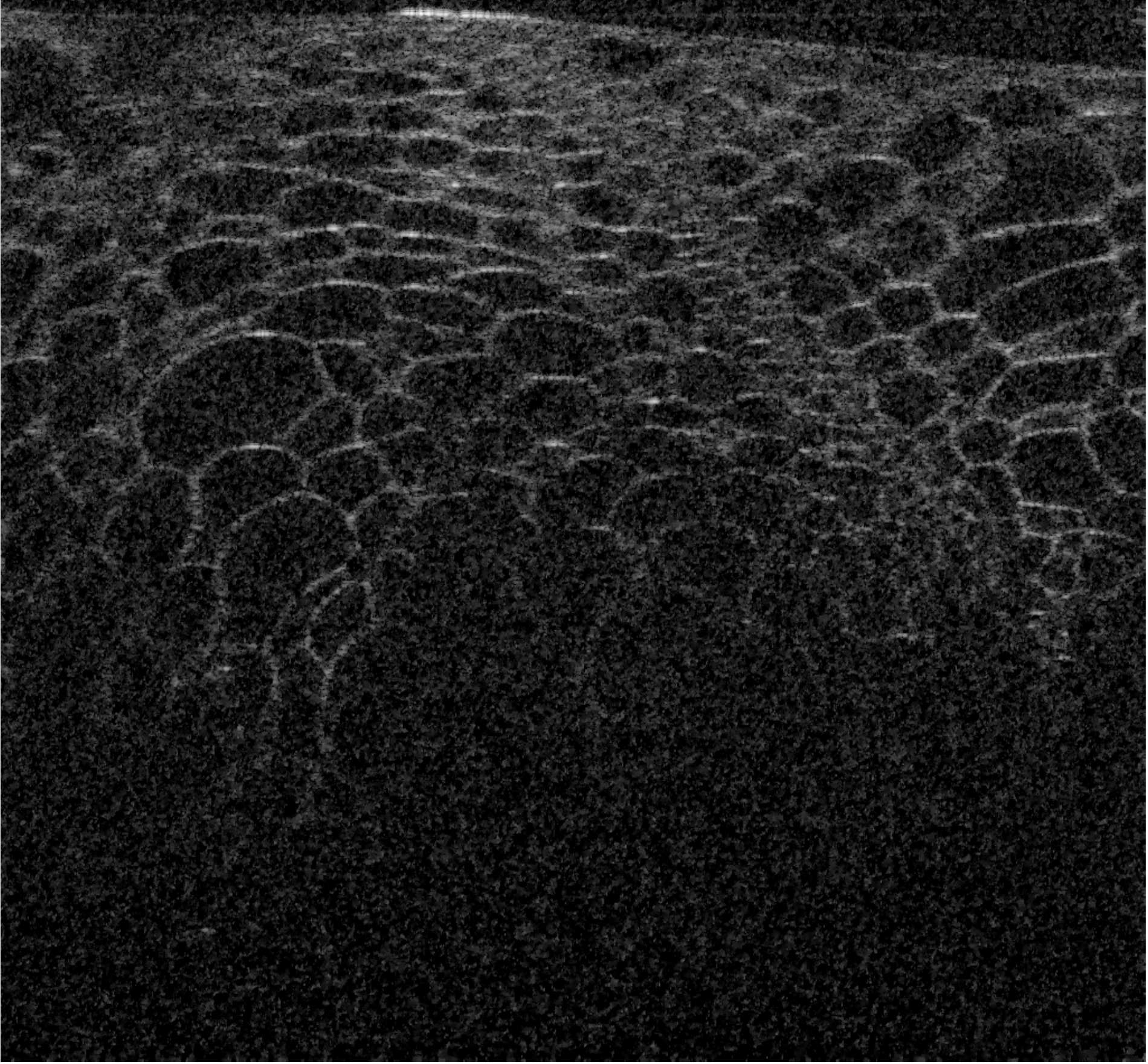}&
			\includegraphics[height=4cm]{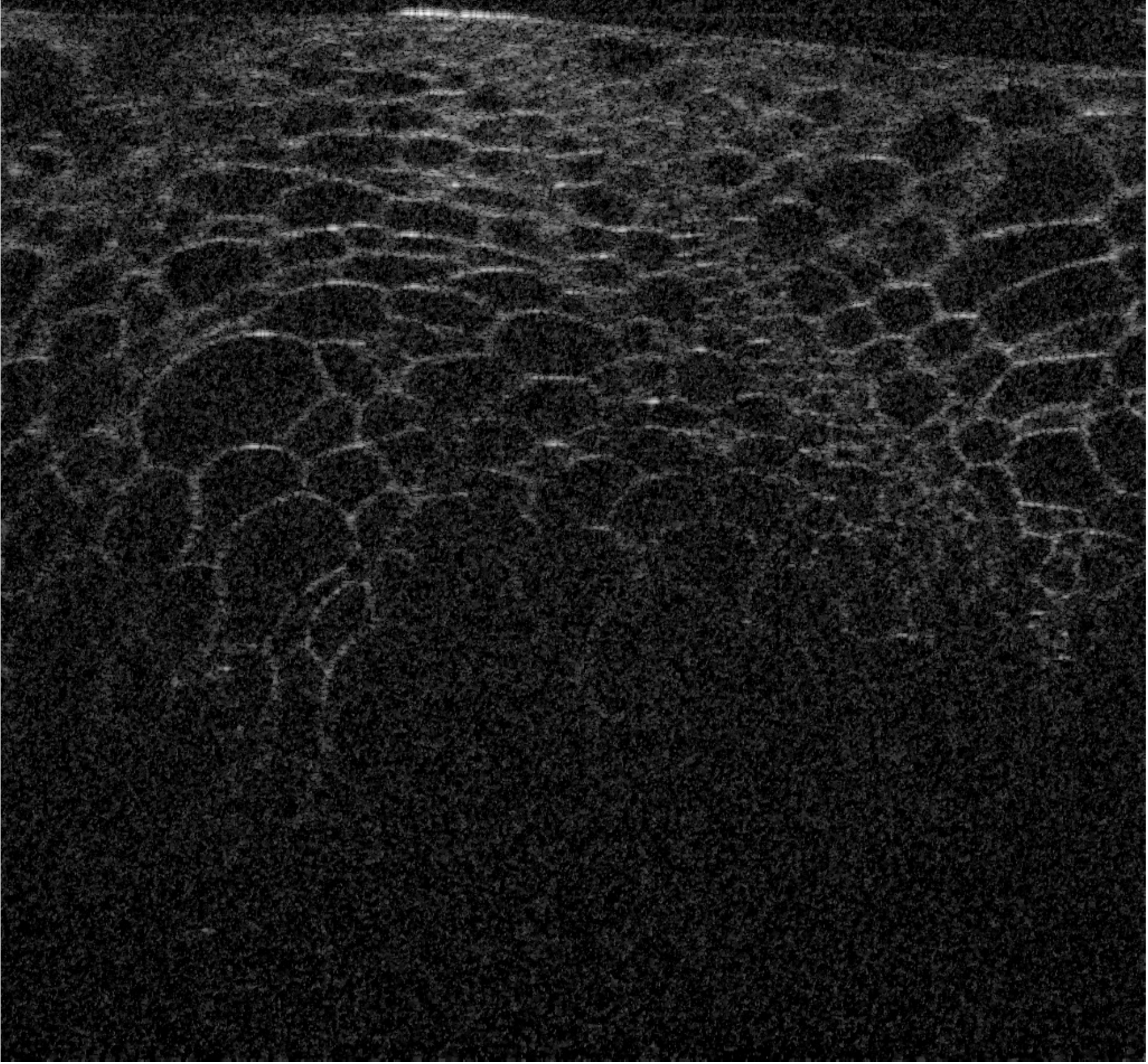}\\
			\hline
			\rotatebox{90}{equispaced}&
			\includegraphics[height=4cm]{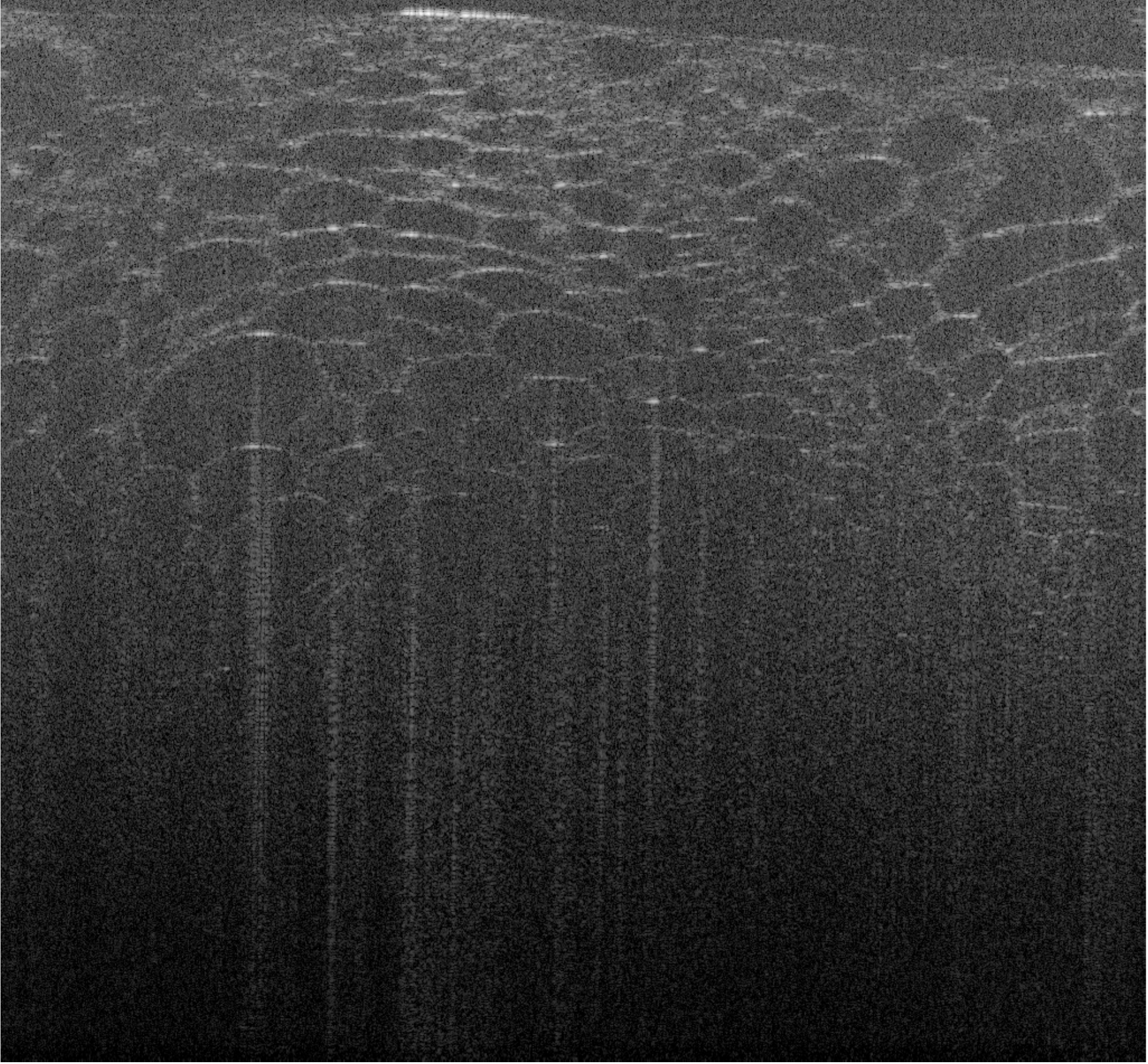}&
			\includegraphics[height=4cm]{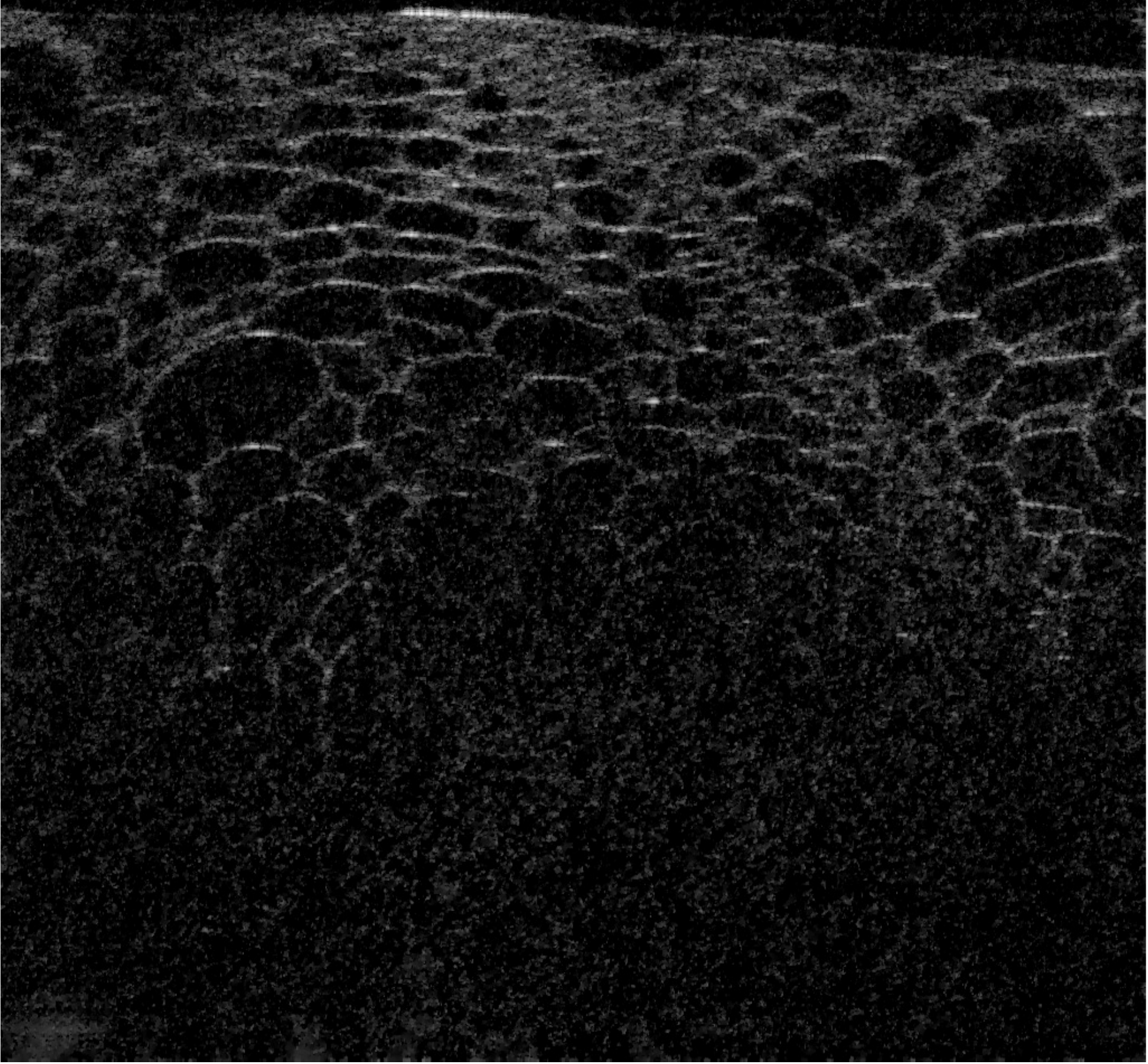}&
			\includegraphics[height=4cm]{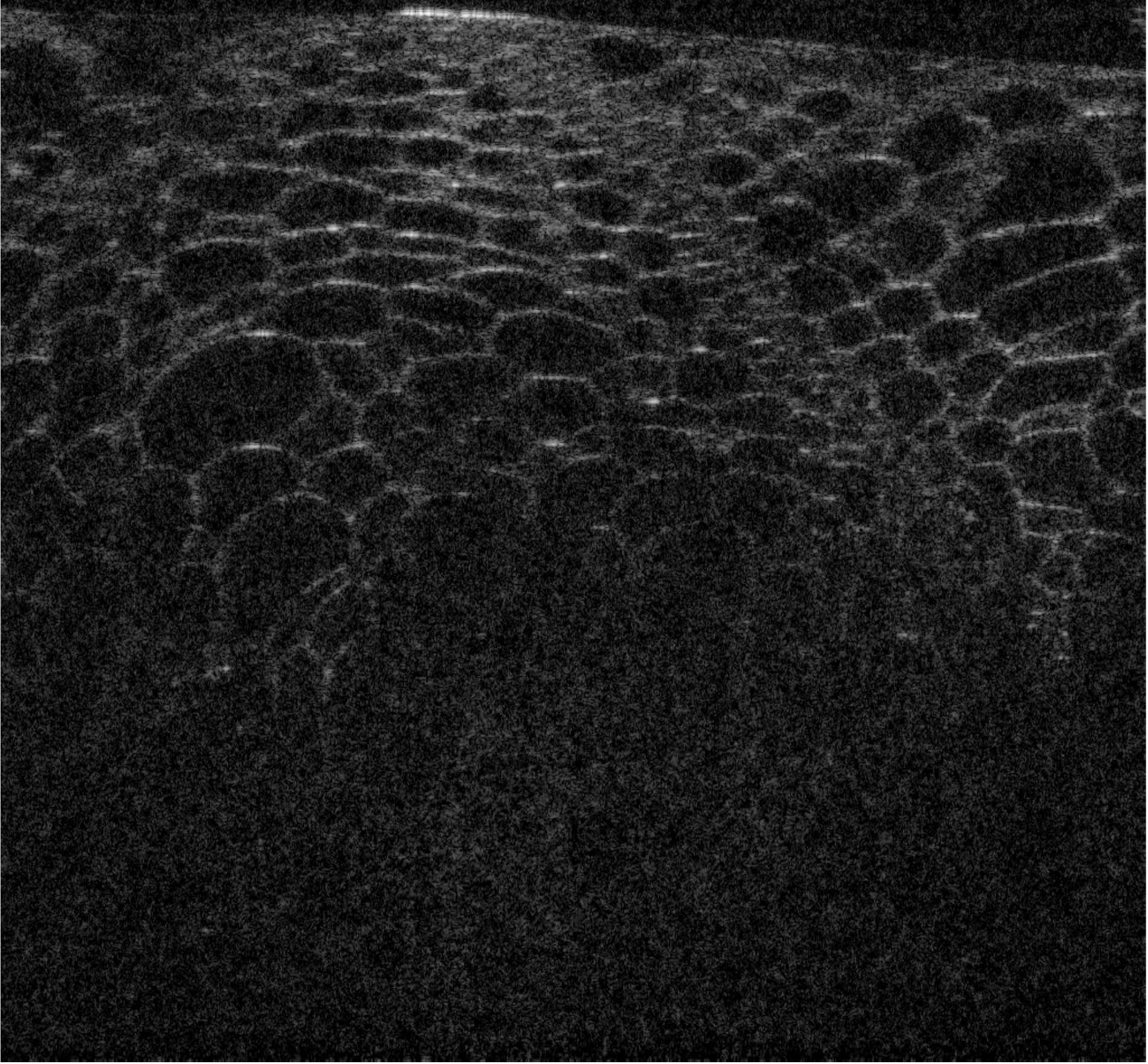}\\
			\hline
			\rotatebox{90}{partial}&
			\includegraphics[height=4cm]{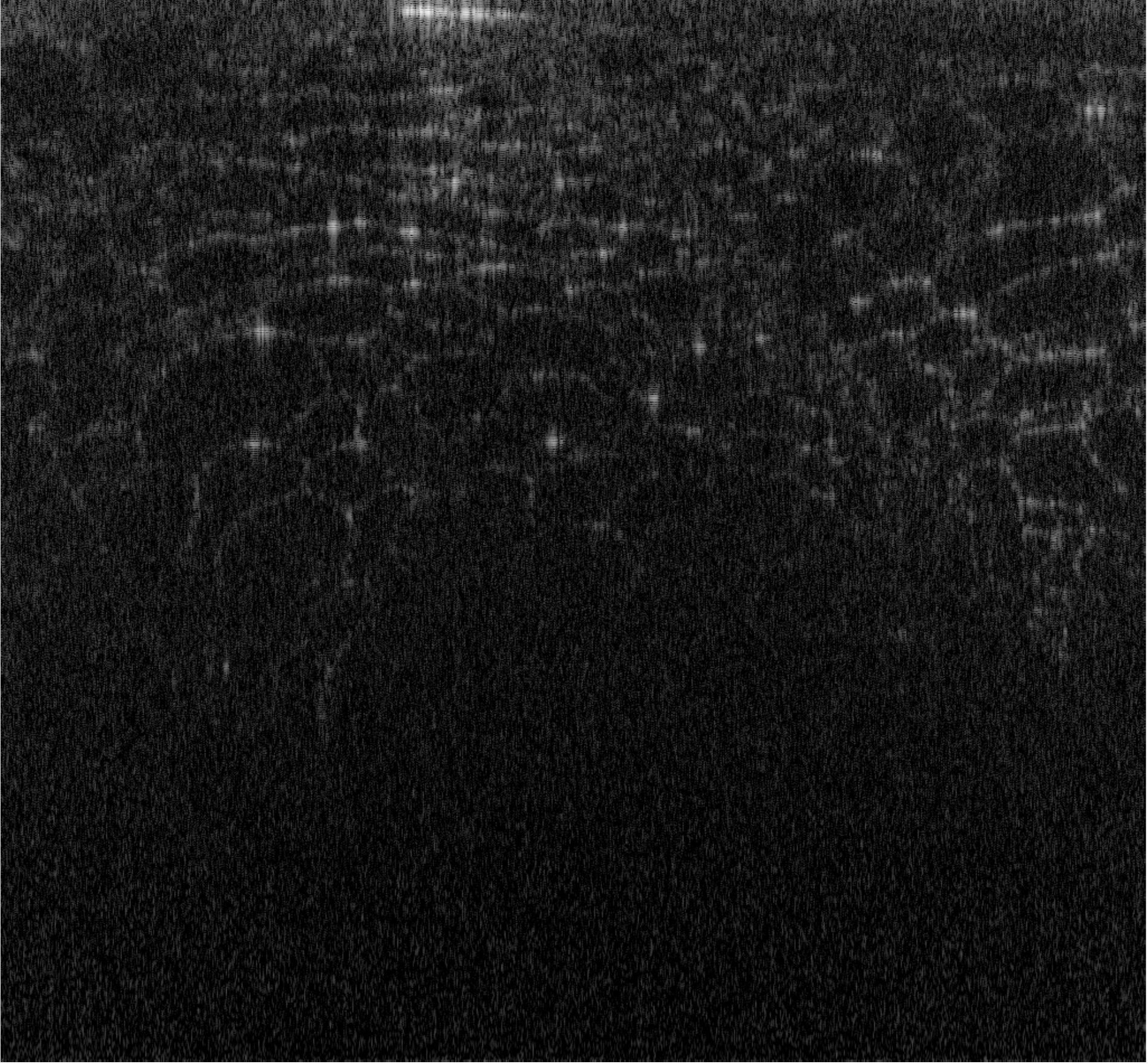}&
			\includegraphics[height=4cm]{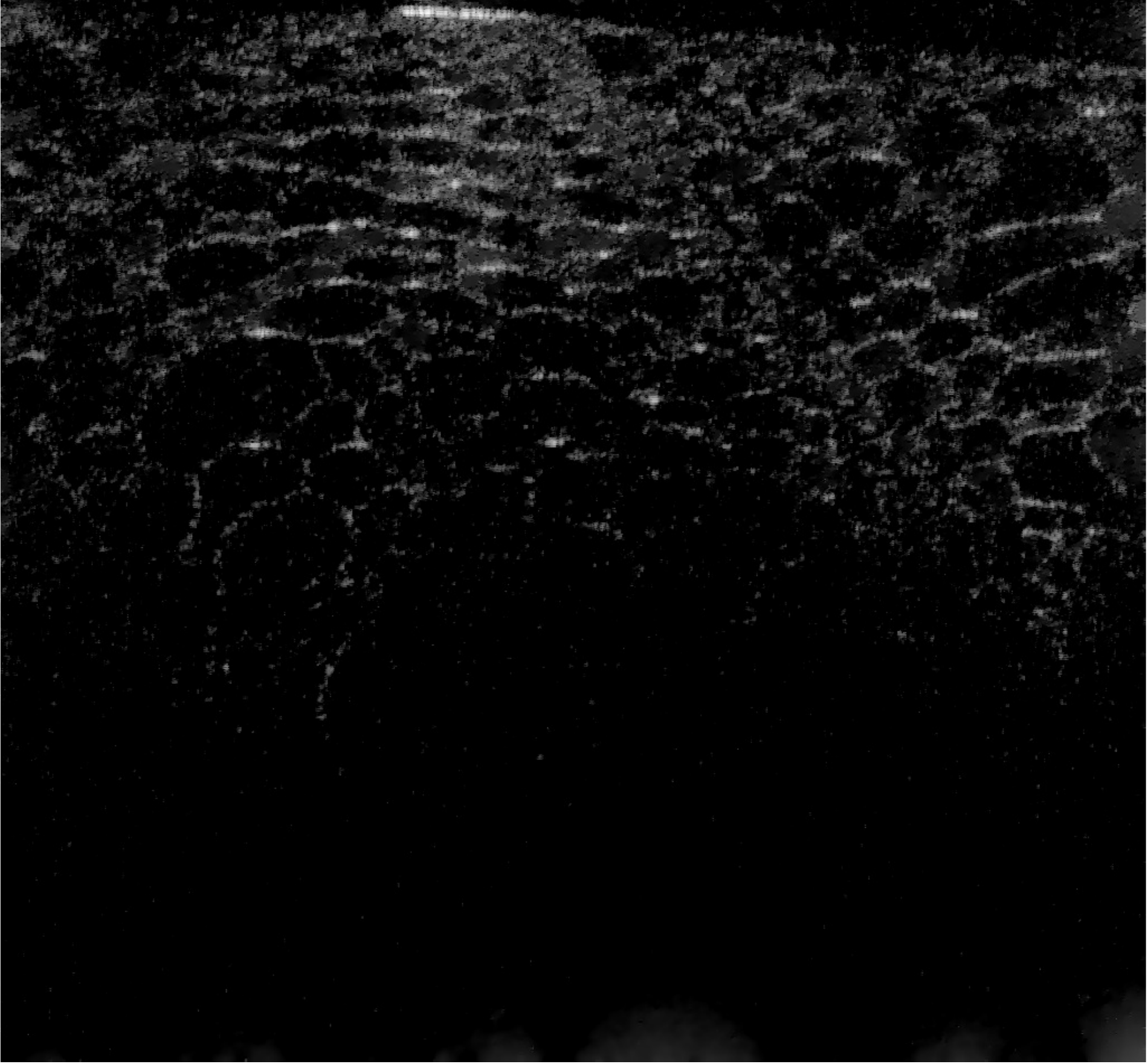}&
			\includegraphics[height=4cm]{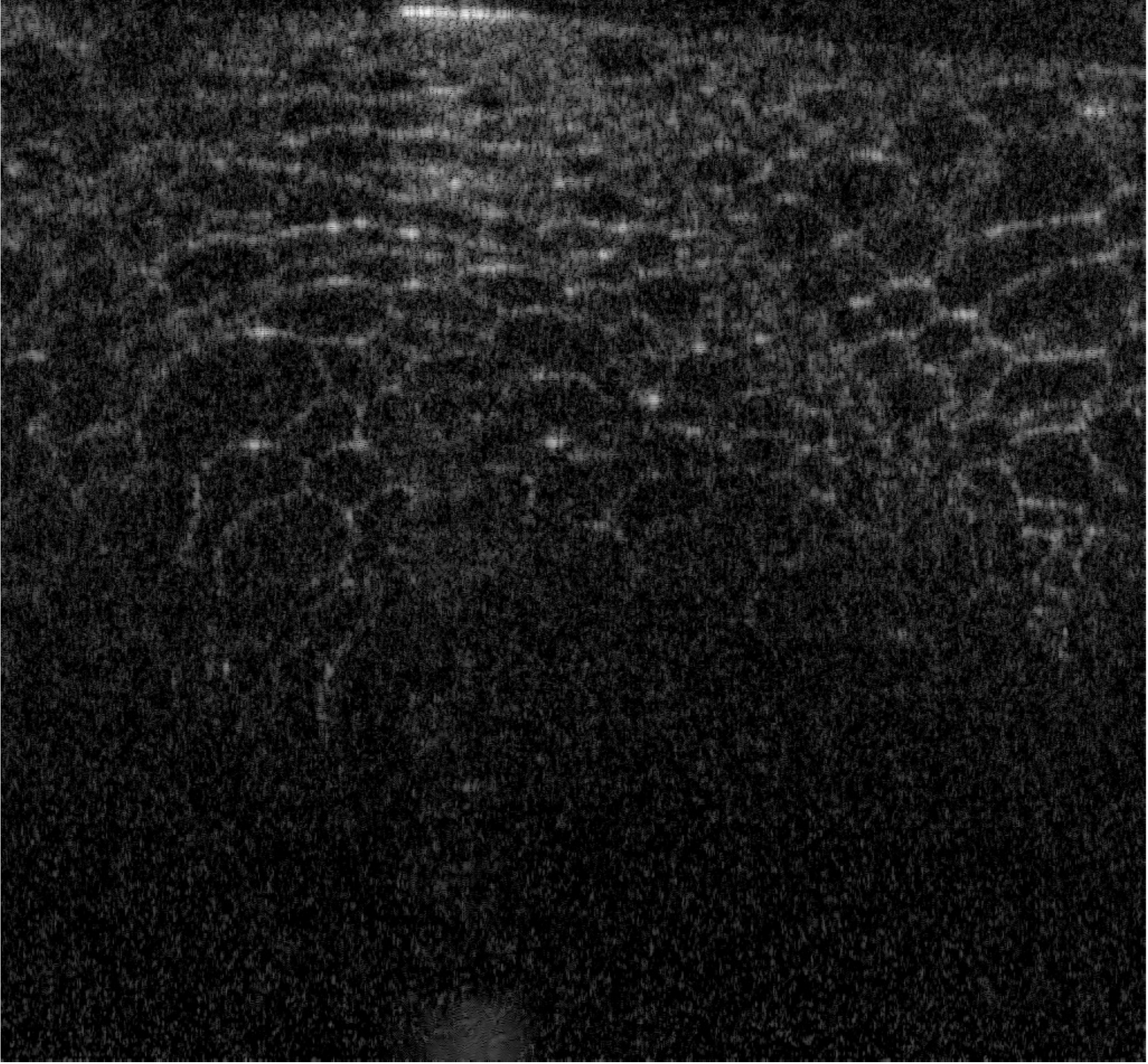}\\
			\hline
		\end{tabular}
	\end{center}
	\caption[example] 
	{\label{fig:cucumber} Visual results of cucumber tissue reconstructions from various sub-sampling schemes.}
\end{figure}

Visualizations of cross sectional reconstructions from the sub-sampled measurements are shown in Figure~\ref{fig:beads} and Figure~\ref{fig:cucumber} for the beaded gel and cucumber tissue samples respectively. Quantitative results of the NCC are then shown in Table~\ref{tab:results}.

From Figure~\ref{fig:beads}, it appears that enhancement of the image structure is improved in each instance of MBIR. Random sub-sampling results in the best preservation of structure, followed by equispaced, and finally partial, which exhibits strong artifacts in each case. TV regularization appears to reasonably robust, but exhibits some block-like image artifacts, whilst DT-CWT has better preservation of speckle structure.

The cucumber tissues results in Figure~\ref{fig:cucumber} follow a similar trend to the beaded phantom, with very good preservation of structure for random and equispaced sub-sampling for both MBIR techniques. Again, partial measurements are significantly worse, with most of the structure degraded in the interpolated ISAM image.

\begin{table}
	\centering
	\begin{tabular}{c|c|c|c|c}
		specimen & fully sampled & IFFT & dispersion corrected & ISAM \\
		\hline
		beaded gel & - & 0.285 & 0.954 & \textbf{1} (reference) \\
		cucumber & - & 0.278 & 0.687 & \textbf{1} (reference) \\
		\hline
		&sub-sampled & interpolated ISAM & MBIR TV & MBIR DT-CWT \\
		\hline
		beaded gel &random & 0.973 & 0.984 & \textbf{0.986} \\
		&equispaced & 0.956 & \textbf{0.984} & 0.981 \\
		&partial & 0.245 & \textbf{0.824} & 0.775 \\
		\hline
		cucumber &random & 0.890 & \textbf{0.925} & 0.923 \\
		&equispaced & 0.877 & \textbf{0.912} & 0.909 \\
		&partial & 0.164 & \textbf{0.697} & 0.666 \\
		\hline
	\end{tabular}
	\caption{\label{tab:results} Quantitative results from reconstructions. All results are normalized cross-correlation (NCC) as in (\ref{equ:ncc}) against fully sampled ISAM image.}
\end{table}

The quantitative results in Table~\ref{tab:results} exhibit a powerful result, which is that for almost every case, MBIR from half the measurements outperforms fully sampled dispersion correction. This implies that without imposing the ISAM model, one can expect a larger loss in quality than with less measurements.

Between the sub-sampling schemes, agreeing our discussion on visual results, equispaced and random sampling perform similarly, with partial detector having poor performance across the board. However, it is noteworthy that MBIR is able to have such a drastic improvement over interpolated ISAM in the partial case.

Between regularization methods, there is generally little difference in quantitative performance, with TV having slightly higher NCC in most cases. In practice, one should consider the observed preservation in speckle structure with DT-CWT, which may be valuable in some applications such as displacement tracking.

\section{CONCLUSIONS}
We have derived a MBIR framework for SD-OCT, including dispersion compensation and ISAM model, and evaluated various sub-sampling schemes and regularization functions. We found that from only half the measurements, we are able to achieve very good structural reconstruction. Random sub-sampling is well performing, but less practical than equispaced, which offers very similar results. These results imply options either to realize SD-OCT with cheaper hardware, or extend the range or resolution of existing systems. Work to implement real-time MBIR is ongoing.
 \section*{ACKNOWLEDGMENTS}       

The authors sincerely thank Graham Anderson from the University of Edinburgh, for assistance creating the beaded gel phantom.
This work was supported by the UK Engineering and Physical Sciences Research Council (EPSRC) MechAScan project: EP/P031250/1.
\bibliographystyle{apalike}
\bibliography{oct_mbir}
\end{document}